\documentclass[preprint,review,3p,times,a4paper]{elsarticle}
\usepackage[linesnumbered,ruled,vlined, ruled,vlined]{algorithm2e}
\usepackage{amsmath}
\usepackage{lineno}
\usepackage{pgfplots}
\pgfplotsset{compat=1.18}
\usepackage{subcaption}
\usepackage{tikz}
\usepackage{xcolor}
\usepgfplotslibrary{fillbetween}

\def\ie{i.e., }

\newcommand{\kh}[1] {\textcolor{black}{#1}}    
\journal{International Journal of Heat and Mass Transfer}
\begin{document}
\begin{frontmatter}
\title{Deep Operator Learning-based Surrogate Models with Uncertainty Quantification for Optimizing Internal Cooling Channel Rib Profiles}
\author[label2]{Izzet Sahin}
\author[label3]{Christian Moya}
\author[label2]{Amirhossein Mollaali}
\author[label2,label3]{Guang Lin\corref{core}}
	\ead{guanglin@purdue.edu}
    \cortext[core]{corresponding author}
\author[label2]{Guillermo Paniagua}

 \affiliation[label2]{organization={Purdue University},
             addressline={School of Mechanical Engineering},
             city={West Lafayette},
             postcode={47906},
             state={IN},
             country={USA}}
 \affiliation[label3]{organization={Purdue University},
			 addressline={Department of Mathematics},
		 	 city={West Lafayette},
	 		 postcode={47906},
			 state={IN},
			 country={USA}}
\begin{abstract}
This paper designs surrogate models with uncertainty quantification capabilities to improve the thermal performance of rib-turbulated internal cooling channels effectively. To construct the surrogate, we use the deep operator network~(DeepONet) framework, a novel class of neural networks designed to approximate mappings between infinite-dimensional spaces using relatively small datasets. The proposed DeepONet takes an arbitrary continuous rib geometry with control points as input and outputs continuous detailed information about the distribution of pressure and heat transfer around the profiled ribs. The datasets needed to train and test the proposed DeepONet framework were obtained by simulating a 2D rib-roughened internal cooling channel. To accomplish this, we continuously modified the input rib geometry by adjusting the control points according to a simple random distribution with constraints, rather than following a predefined path or sampling method. The studied channel has a hydraulic diameter, $D_{h}$, of 66.7 mm, and a length-to-hydraulic diameter ratio, $L/D_{h}$, of 10. The ratio of rib center height to hydraulic diameter ($e/D_{h}$), which was not changed during the rib profile update, was maintained at a constant value of 0.048. The ribs were placed in the channel with a pitch-to-height ratio ($P/e$) of 10. In addition, we provide the proposed surrogates with effective uncertainty quantification capabilities. This is achieved by converting the DeepONet framework into a Bayesian DeepONet (B-DeepONet). B-DeepONet samples from the posterior distribution of DeepONet parameters using the novel framework of stochastic gradient replica-exchange MCMC. Finally, we demonstrate the performance of the proposed DeepONet-based surrogate models with uncertainty quantification by incorporating them into a constrained, gradient-free optimization problem that enhances the thermal performance of the rib-turbulated internal cooling channel.

\end{abstract}
\end{frontmatter}

\section{Introduction} \label{sec:Introduction}
To improve heat transfer performance in advanced gas turbine internal cooling channels, turbulators such as ribs and pin-fins are used in mid-chord and trailing edge regions \cite{HanGT,2004Wright1,2021Sahin2}. Ribs, which are usually placed in the mid-chord internal cooling channels, disturb the mainstream flow near the ribbed surface, creating turbulated and rib-induced secondary flow, and increasing heat transfer rates. However, ribs also increase channel flow resistance and bring pressure losses. To investigate the effect of mainstream flow and rib-induced secondary flow interactions on heat transfer rate and pressure drop, rib parameters such as rib spacing, orientation, and profile have been studied numerically and experimentally in stationary and rotating internal cooling channels \cite{HanGT,2012Liu1,2022Luke1}.

In order to understand the effects of rib spacing on heat transfer and pressure drop, the rib pitch-to-height ratio ($P/e$) was varied from 5 to 20 while maintaining a constant rib height-to-hydraulic diameter ratio ($e/D_{h}$) for various rib orientations. Flow behavior, such as circulation, separation, and reattachment locations, and their effects were previously characterized \cite{1978Han1}. It was found that increasing the $P/e$ ratio beyond 10 weakened the rib effects on flow separations and reattachments, resulting in reduced heat transfer. On the other hand, reducing the $P/e$ ratio below 10 led to a low heat transfer rate due to the trapped flow circulation between adjacent ribs. Therefore, it was recommended to use a $P/e$ ratio of around 10 to obtain relatively high thermal performance from internal cooling channels, based on evaluations of both heat transfer and pressure drop performances \cite{2016Han1}.

In \cite{1988HAN183}, the effect of rib orientation and rib-induced secondary flow development on channel heat transfer and pressure drop behavior was studied while keeping the rib pitch-to-height ratio ($P/e$) constant. It was observed that when rib orientation angles deviate from 90$^\circ$ in orthogonal cases, the rib-induced secondary flow characteristics and direction in the channel change significantly. In \cite{2015CHUNG357} and \cite{2020Sahin1}, the authors outlined high and low heat transfer regions and their effect on rib-roughened surfaces and channel side walls, both numerically and experimentally. Furthermore, to eliminate reduced heat transfer rate regions on rib-roughened surfaces, intersecting, broken, v-shaped, and w-shaped ribs were investigated for single and multi-pass channels under stationary and rotating conditions in \cite{2015CHUNG357} and \cite{2004Wright1}.

In terms of rib profiles, research has focused on reducing high-pressure penalties and promoting heat transfer in lower regions near the ribs, where flow stagnation and circulation occur. Pioneer studies have examined continuous and broken wedge-shaped, delta-shaped \cite{1994HAN11}, square, rectangular, and profiled ribs \cite{1998Chandra1}. These studies found that delta-shaped ribs, as opposed to wedge-shaped ones, have a positive effect on heat and pressure drop performance. Delta-shaped ribs mitigate flow circulation at downstream regions of ribs, which further improves channel pressure behavior. Regarding rib profiles, channels with square ribs provide higher heat transfer performance and pressure drop than those with profiled ribs due to the flow impingement effect on the perpendicular wall of square ribs.

The studies mentioned above demonstrate that turbulators, such as ribs, are essential for maintaining the operation of gas turbine internal cooling channels. Therefore, researchers have focused on determining the optimal range of rib parameters to improve the efficiency of these channels using numerical and experimental approaches. However, due to limitations in time and resources, researchers investigated only a limited number of samples. Additionally, in the past, manufacturing capabilities restricted the geometry of profiled ribs. Nowadays, with the development of advanced additive manufacturing techniques, any rib profile may be considered in turbine blade internal cooling channels. 

Furthermore, in addition to numerical and experimental studies, machine learning surrogate models have been proposed to optimize the parameters of internal cooling channels and improve their thermal performance. For example, Verstraete et al.~\cite{2013Verstraete1} used Artificial Neural Network (ANN) to optimize the U-bend channel by updating its geometry with control points. At around 37\%, pressure recovery was observed in the two-pass internal cooling channel that utilized the optimized U-bend region. In~\cite{2022Wang1}, the authors compared several conventional and deep learning-based neural network models for the U-bend regions, and deep learning-based models were shown to have higher response accuracy and better ability to capture heat transfer and pressure distribution in the channel. Keramati et al.~\cite{2022Keramati1} investigated shape optimization for heat exchanger geometry in a 2D flow domain using Deep Neural Networks and finite element numerical solver, achieving over 30\% heat transfer enhancement and 60\% pressure recovery. Polat and Cadirci~ \cite{2022Polat1} conducted parametric optimization of diamond-shaped fin geometry in a single pass micro-channel using the ANN model. They enhanced heat transfer without a significant pressure drop by optimizing the diamond fin geometry while maintaining the constant rhomboidal pin-fin area for each design point. Rebassa \cite{Rebassa2023120} used a multi-objective approach for fin shape optimization to enhance heat transfer and reduce pressure drop for heat exchangers used in the bypass duct of turbofan engines. They obtained around 7.5\% heat transfer enhancement and 9.1\% pressure drop reduction in the optimized geometry. Finally, in~\cite{2016Kim1}, the authors optimized the boot-shaped rib profile in a single pass internal cooling channel using numerical methods, and comparatively tested two surrogate models based on response surface approximation (RSA) and Kriging methods. They illustrated the rib geometry effect on heat transfer and friction factor for several boot-shaped rib profiles. 

The surrogate models used in the aforementioned optimization studies require a large amount of high-fidelity data to ensure a small approximation error. Additionally, they can only provide discrete information about the internal cooling channel.  However, field information, mainly collected through experimental and numerical studies in a cost-effective way, is needed to optimize channel parameters properly. Therefore, in this paper, we use the Deep Operator Network (DeepONet), proposed in \cite{2021Lulu1} as a method to approximate nonlinear operators, which are mappings between infinite-dimensional spaces, to obtain a surrogate model that can reduce rib optimization costs.

DeepONet is based on the universal approximation theorem of nonlinear operators~\cite{392253} and it can effectively approximate nonlinear operators that arise in complex dynamical systems using relatively small datasets. As a result, DeepONet has been successfully applied to model various applications. For example, it accurately captured the flow parameters of hypersonic flow conditions in \cite{2021LuLu2}, even outside the training range of Mach conditions. In \cite{2022Lin21}, DeepONet learned the dynamic response of nonlinear control systems. In~\cite{MOYA2023166}, we proposed a novel DeepONet-based approach for reliable forecasting of post-fault trajectories of power grids. Additionally, graph-based DeepONet was introduced in~\cite{sun2022deepgraphonet} and effectively forecasted the dynamic response of networked dynamical systems. Moreover, multi-fidelity DeepONet was successfully used to further reduce the high-fidelity dataset while maintaining its high accuracy, as demonstrated in \cite{2022Lulu1}. 

Although DeepONet is effective, it may struggle when trained with small and noisy data. To address this issue, we developed a framework called Bayesian DeepONet (B-DeepONet) in~\cite{lin2023b}, which achieves high-accuracy results when trained with noisy data. B-DeepONet was trained using the stochastic gradient replica exchange Langevin diffusion method, which we studied and developed in~\cite{deng2020non,lin2023b}, instead of the classic stochastic gradient-based training. B-DeepONet has been used for several applications and has obtained highly accurate results. Moreover, B-DeepONet can provide reliable predictions via confidence intervals and reduce training costs using accelerated training regimes~\cite{lin2023b}.

This paper aims to capture the heat transfer and pressure drop behavior of any randomly chosen rib profile in a channel and optimize it for orthogonal-oriented rib-roughened internal cooling channels. To accomplish this, we leverage the power of the DeepONet and B-DeepONet surrogate models trained with data obtained using numerical methods.

The main contributions of this paper are:

\begin{itemize}
    \item We design and train (Section~\ref{sec:DeepONet}) a DeepONet-based surrogate model that takes an arbitrary continuous rib geometry with control points as input and outputs continuous detailed information about the distribution of pressure and heat transfer around the profiled ribs. DeepONet allows for a detailed analysis of the impact of near-rib flow circulation on pressure and the Nusselt number, for any desired rib profiles, without the need for numerical simulation.
    \item To address scenarios where the training data is scarce and noisy, we construct (Section~\ref{sec:B-DeepONet}) a Bayesian DeepONet (B-DeepONet) as a surrogate model. B-DeepONet enables quantifying the uncertainty for the mapping between arbitrary rib profiles and the distribution of pressure and heat transfer around the profiled ribs.
    \item To demonstrate the effectiveness of DeepONet and B-DeepONet, we have designed (Section~\ref{sec:optimization}) a constrained gradient-free optimization framework that employs either DeepONet or B-DeepONet as surrogates for enhancing internal cooling channels. Specifically, we will illustrate that this framework can (i) rapidly generate rib profiles, and (ii) prevent adversarial rib designs, which represent a significant challenge for optimization procedures based on machine learning.
\end{itemize}


\section{Problem Formulation} \label{sec:problem-formulation}
\begin{figure}[!b] 
	\centering{\includegraphics[width=3.0in]{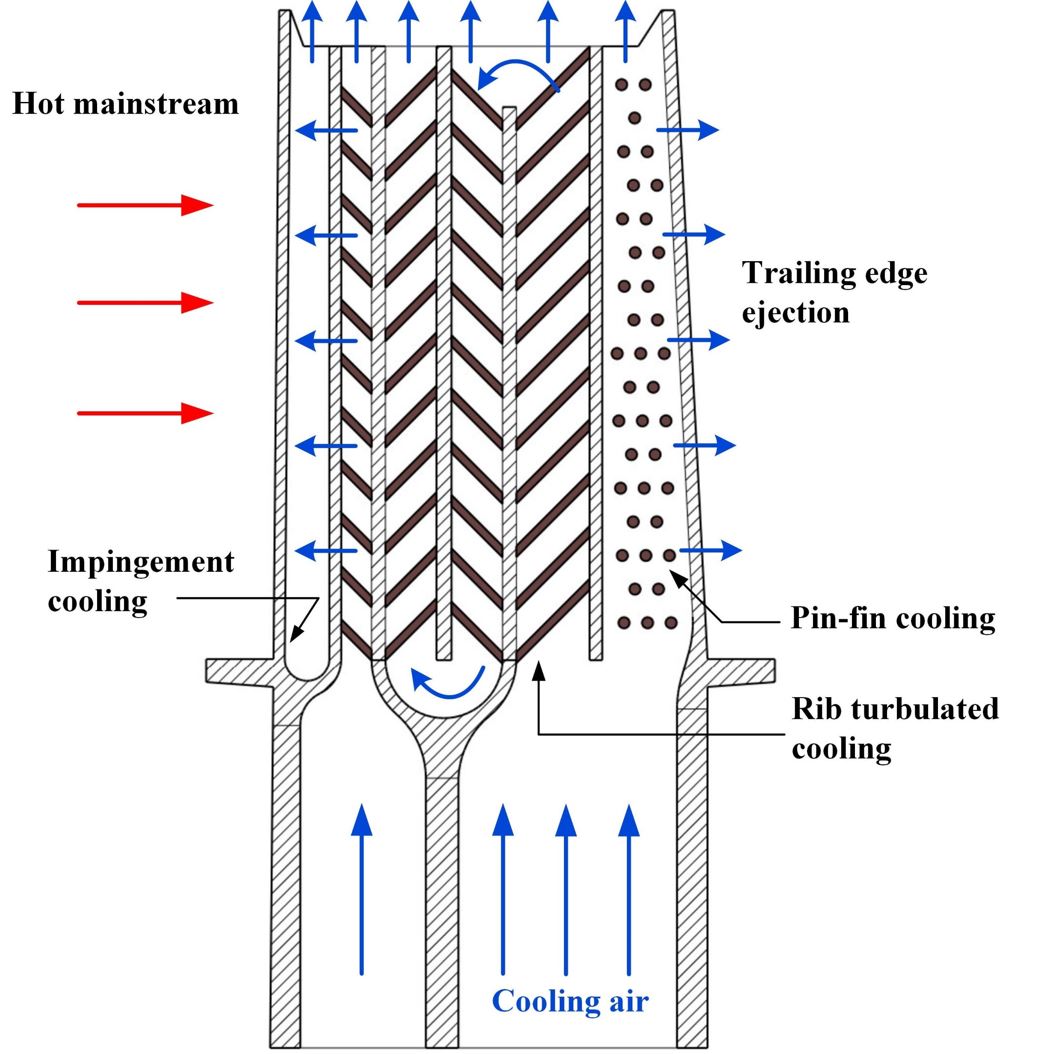}}
	\caption{Internal cooling techniques in an advanced turbine blade \cite{2006Han1}.}
	\label{fig:cooling-schematic}
\end{figure}

To enhance the internal cooling performance of modern gas turbine blades and vanes, various techniques such as jet impingement, rib turbulated, and pin-fin cooling methods are applied, as shown in Figure~\ref{fig:cooling-schematic}. Rib turbulators are particularly effective in enhancing heat transfer in the mid-chord of blades, as depicted in Figure~\ref{fig:cooling-schematic}. Thus, researchers have experimentally and numerically investigated the effects of rib parameters, specifically in high aspect ratio (AR = $W/H > 1$), rectangular cross-sectional, stationary, and rotating channels, to detail the thermal and hydraulic behavior of internal cooling channels.

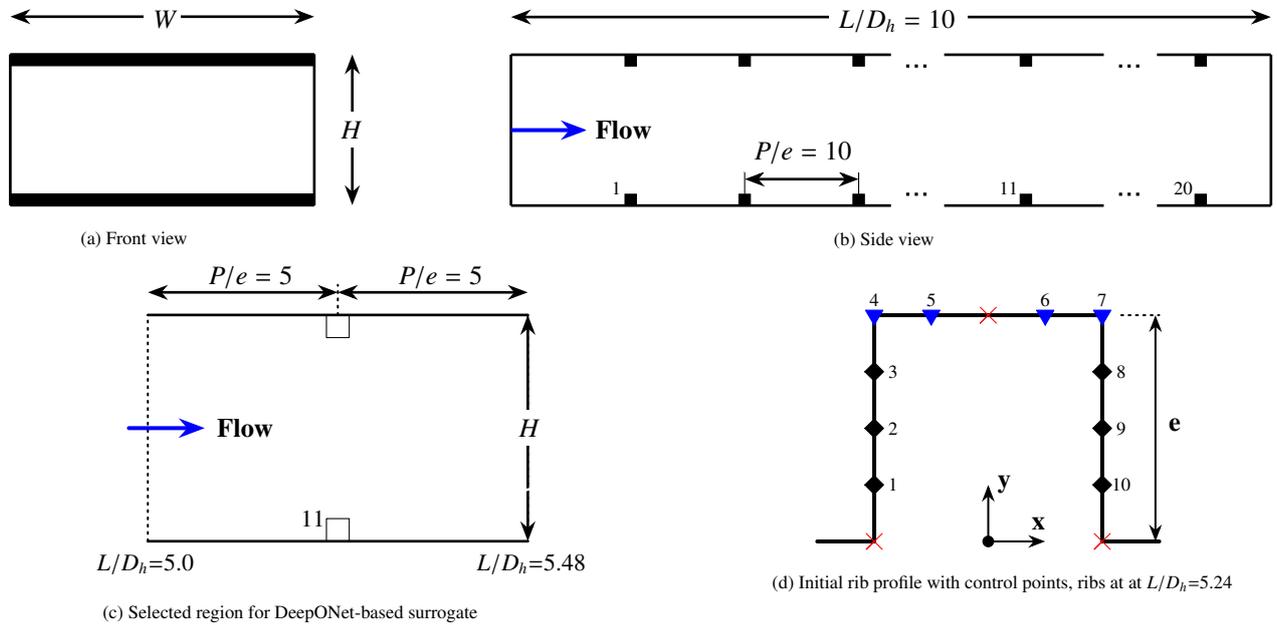
\begin{figure}[!b] 
	\centering
	\usetikzlibrary {arrows.meta}
	\begin{subfigure}[bl]{0.2\textwidth}
		\begin{tikzpicture}	
			\draw [line width=1pt] (0,0)-- (0,2);
			\draw [line width=1pt] (0,2)-- (4,2);
			\draw [line width=1pt] (4,2)-- (4,0);
			\draw [line width=1pt] (4,0)-- (0,0);
			\filldraw[draw=black,fill=black] (0,0) rectangle (4,0.15);
			\filldraw[draw=black,fill=black] (0,1.85) rectangle (4,2);
			\draw [Stealth-Stealth,line width=1pt,color=black] (4.5,0)-- (4.5,2);
			\node[text width=7.0,fill=white] at (4.47, 1.0) {$H$};		
			\draw [Stealth-Stealth,line width=1pt,color=black] (0,2.5)-- (4,2.5);
			\node[text width=7.0,fill=white] at (2, 2.47) {$W$};	
		\end{tikzpicture}	
		\caption{Front view}
		\label{Channelfront}	
	\end{subfigure}	
	\hfill
	\begin{subfigure}[br]{0.6\textwidth}
		\begin{tikzpicture}
			\draw [line width=1pt] (0,0)-- (0,2);
			\draw [line width=1pt] (0,2)-- (5,2);
			\draw[draw=black, line width=1.2pt, dotted] (5.20,1.85) -- (5.50,1.85);
			\draw [line width=1pt] (5,0)-- (0,0);
			\draw[draw=black, line width=1.2pt, dotted] (5.20,0.15) -- (5.50,0.15);
			\draw [line width=1pt] (5.70,2)-- (7.80,2);
			\draw[draw=black, line width=1.2pt, dotted] (8.0,1.85) -- (8.30,1.85);
			\draw [line width=1pt] (8.5,2)-- (10,2);
			\draw [line width=1pt] (10,2)-- (10,0);
			\draw [line width=1pt] (10,0)-- (8.5,0);
			\draw[draw=black, line width=1.2pt, dotted] (8.0,0.15) -- (8.30,0.15);
			\draw [line width=1pt] (5.70,0)-- (7.80,0);
			\filldraw[draw=black,fill=black] (1.5,0) rectangle (1.65,0.15);
			\filldraw[draw=black,fill=black] (3,0) rectangle (3.15,0.15);
			\filldraw[draw=black,fill=black] (4.5,0) rectangle (4.65,0.15);
			\filldraw[draw=black,fill=black] (6.7,0) rectangle (6.85,0.15);
			\filldraw[draw=black,fill=black] (9,0) rectangle (9.15,0.15);
			\filldraw[draw=black,fill=black] (1.5,1.85) rectangle (1.65,2);
			\filldraw[draw=black,fill=black] (3,1.85) rectangle (3.15,2);
			\filldraw[draw=black,fill=black] (4.5,1.85) rectangle (4.65,2);
			\filldraw[draw=black,fill=black] (6.7,1.85) rectangle (6.85,2);
			\filldraw[draw=black,fill=black] (9,1.85) rectangle (9.15,2);
			\draw [Stealth-Stealth,line width=1pt,color=black] (0,2.5)-- (10,2.5);
			\node[text width=45.0,fill=white] at (5.1, 2.45) {$L/D_h=10$};
			\draw [-Stealth,line width=1.5pt,color=blue] (0,1)-- (1,1);
			\node[text width=5.0,fill=white] at (1.2, 1) {\textbf{Flow}};
			\draw[draw=black] (3.075,0.15) -- (3.075,0.45);
			\draw[draw=black] (4.575,0.15) -- (4.575,0.45);
			\draw [Stealth-Stealth,line width=1pt,color=black] (3.075,0.35)-- (4.575,0.35);
			\node[text width=40.0,fill=white] at (3.9, 0.7) {$P/e=10$};
			\begin{scriptsize}
				\node[text width=10.0] at (1.5, 0.23) {$1$};
				\node[text width=10.0] at (6.60, 0.23) {$11$};
				\node[text width=10.0] at (8.90, 0.23) {$20$};
			\end{scriptsize}
		\end{tikzpicture}
		\caption{Side view}
		\label{subfig:channel-side}				
	\end{subfigure}	
	\hfill
	\begin{subfigure}[br]{0.45\textwidth}
		\centering
		\begin{tikzpicture}[remember picture, line cap=round, line join=round,scale=1]
		\draw node[anchor=north west, inner sep=0](model){};
		\node[text width=40.0,fill=white] at (3, 3.5) {$P/e=5$};
		\draw [Stealth-Stealth,line width=1pt,color=black,] (1.5,3.3) -- (4,3.3);
		\filldraw[draw=black,fill=white] (3.85,3.0) rectangle (4.15,2.7);
		\node[text width=40.0,fill=white] at (5.5, 3.5) {$P/e=5$};
		\filldraw[draw=black,fill=white] (3.85,0.0) rectangle (4.15,0.3);
		\draw [Stealth-Stealth,line width=1pt,color=black,] (4,3.3) -- (6.5,3.3);
		\draw[draw=black, line width =0.75pt, dotted] (4,3.0)-- (4,3.4);	
		\draw [draw=black,line width=0.75pt, dotted] (1.5,0)-- (1.5,3);
		\draw [draw=black,line width=1pt] (1.5,3)-- (6.5,3);
		\draw[draw=black, line width=0.75pt, dotted] (6.5,3) -- (6.5,0);
		\draw [draw=black,line width=1pt] (6.5,0)-- (1.5,0);
		\begin{scriptsize}
			\node[text width=5.0, font={\small}] at (3.6, 0.3) {$11$};
            \node[text width=10.0, font={\small}] at (1.0, -0.3) {$L/D_h$=5.0};
            \node[text width=10.0, font={\small}] at (6.0, -0.3) {$L/D_h$=5.48};
		\end{scriptsize}
		\draw [-Stealth,line width=1.5pt,color=blue] (1.25,1.5)-- (2.25,1.5);
		\node[text width=5.0,fill=white] at (2.5, 1.5) {\textbf{Flow}};
		\draw [Stealth-Stealth,line width=1pt,color=black] (6.5,0)-- (6.5,3);
		\node[text width=7.0,fill=white] at (6.5, 1.5) {$H$};	
		\end{tikzpicture}
		\caption{Selected region for DeepONet-based surrogate}
		\label{subfig:rectangular2D}
	\end{subfigure}
	\hfill	
	\begin{subfigure}[bl]{0.4\textwidth}
	\centering
	\begin{tikzpicture}[remember picture, line cap=round,line join=round, scale=1]
		\draw node[anchor=north west, inner sep=0](replica){};
		\draw [line width=1.5pt] (0.0,0.0)-- (0.75,0.0);
		\draw [line width=1.5pt] (0.75,0)-- (0.75,3.0);
		\draw [line width=1.5pt] (0.75,3.0)-- (3.75,3.0);
		\draw [line width=0.75pt,dotted] (4.0,3)-- (4.5,3);
		\draw [line width=1.5pt] (3.75,3.0)-- (3.75,0.0);
		\draw [line width=1.5pt] (3.75,0.0)-- (4.5,0.0);
		\draw [Stealth-Stealth,line width=0.75pt] (4.45,3)-- (4.45,0);
		\draw (4.5,1.75) node[anchor=north west] {\textbf{e}};
		\draw [-Stealth,line width=0.75pt] (2.25,0)-- (2.25,0.75);
		\draw [-Stealth,line width=0.75pt] (2.25,0)-- (3,0);
		\draw (2.7,0.45) node[anchor=north west] {\textbf{x}};
		\draw (2.25,1.0) node[anchor=north west] {\textbf{y}};
		\draw [fill=black] (2.25, 0) circle (2pt);
		
		\begin{scriptsize}
			\draw [color=red] (0.75,0)-- ++(-3pt,-3pt) -- ++(6pt,6pt) ++(-6pt,0) -- ++(6pt,-6pt);
			\draw [color=red] (2.25,3)-- ++(-3pt,-3pt) -- ++(6pt,6pt) ++(-6pt,0) -- ++(6pt,-6pt);
			\draw [color=red] (3.75,0)-- ++(-3pt,-3pt) -- ++(6pt,6pt) ++(-6pt,0) -- ++(6pt,-6pt);
			\draw [color=blue, fill=blue,shift={(0.75,3)},rotate=180] (0,0) ++(0 pt,3.75pt) -- ++(3.25pt,-5.625pt)--++(-6.5pt,0 pt) -- ++(3.25pt,5.6pt);
			\draw[color=black] (0.75,3.2) node {$4$};
			\draw [color=blue, fill=blue,shift={(1.5,3)},rotate=180] (0,0) ++(0 pt,3.75pt) -- ++(3.25pt,-5.625pt)--++(-6.5pt,0 pt) -- ++(3.25pt,5.6pt);
			\draw[color=black] (1.5,3.2) node {$5$};
			\draw [color=blue,fill=blue,shift={(3,3)},rotate=180] (0,0) ++(0 pt,3.75pt) -- ++(3.25pt,-5.625pt)--++(-6.5pt,0 pt) -- ++(3.25pt,5.625pt);
			\draw[color=black] (3,3.2) node {$6$};	
			\draw [color=blue,fill=blue,shift={(3.75,3)},rotate=180] (0,0) ++(0 pt,3.75pt) -- ++(3.25pt,-5.625pt)--++(-6.5pt,0 pt) -- ++(3.25pt,5.625pt);
			\draw[color=black] (3.75,3.2) node {$7$};		
			
			\draw [fill=black] (0.75,0.75) ++(-3.5pt,0 pt) -- ++(3.5pt,3.5pt)--++(3.5pt,-3.5pt)--++(-3.5pt,-3.5pt)--++(-3.5pt,3.5pt);
			\draw[color=black] (1.0,0.75) node {$1$};
			\draw [fill=black] (0.75,1.5) ++(-3.5pt,0 pt) -- ++(3.5pt,3.5pt)--++(3.5pt,-3.5pt)--++(-3.5pt,-3.5pt)--++(-3.5pt,3.5pt);
			\draw[color=black] (1.0,1.5) node {$2$};	
			\draw [fill=black] (0.75,2.25) ++(-3.5pt,0 pt) -- ++(3.5pt,3.5pt)--++(3.5pt,-3.5pt)--++(-3.5pt,-3.5pt)--++(-3.5pt,3.5pt);
			\draw[color=black] (1.0,2.25) node {$3$};		

			\draw [fill=black] (3.75,2.25) ++(-3.5pt,0 pt) -- ++(3.5pt,3.5pt)--++(3.5pt,-3.5pt)--++(-3.5pt,-3.5pt)--++(-3.5pt,3.5pt);
			\draw[color=black] (4.0,2.25) node {$8$};
			\draw [fill=black] (3.75,1.5) ++(-3.5pt,0 pt) -- ++(3.5pt,3.5pt)--++(3.5pt,-3.5pt)--++(-3.5pt,-3.5pt)--++(-3.5pt,3.5pt);
			\draw[color=black] (4.0,1.5) node {$9$};	
			\draw [fill=black] (3.75,0.75) ++(-3.5pt,0 pt) -- ++(3.5pt,3.5pt)--++(3.5pt,-3.5pt)--++(-3.5pt,-3.5pt)--++(-3.5pt,3.5pt);
			\draw[color=black] (4.0,0.75) node {$10$};		
		\end{scriptsize}
	\end{tikzpicture}
	\caption{Initial rib profile with control points, ribs at at $L/D_h$=5.24}
	\label{subfig:rib-points}
        \end{subfigure}
	\begin{tikzpicture}[overlay, remember picture]
	\draw[-Stealth,line width=1pt,white] ([shift={(4.25, 0.25)}]model.west)--([shift={(0, 1.5)}]replica.east);
	\end{tikzpicture}
    \caption{Schematic views of orthogonal rib roughened internal cooling channel}
    \label{fig:channel-detail}
\end{figure}

This paper uses a rectangular cross-sectional channel model, as shown in Figure~\ref{fig:channel-detail}, which was replicated from~\cite{1988HAN2}. The channel cross-section has a width ($W$) of 100 mm and a height ($H$) of 50 mm, resulting in an aspect ratio of 2:1 ($W/H$) and a hydraulic diameter of 66.7 mm. The channel length is 667 mm, corresponding to a length-to-diameter ratio ($L/D_h$) of 10. The initial rib profile chosen was a square rib with a height ($e$) of 3.2 mm. The rib height to hydraulic diameter ratio is $e/D_h$=0.048. 20 parallel pairs of ribs were placed in an orthogonal manner on the bottom and top surfaces, \ie with a rib orientation angle of $\alpha$ = 90$^\circ$. The spacing between two adjacent ribs was defined based on a pitch-to-height ($P/e$) ratio of 10. As this study focuses on the parametric optimization of the rib profile, and the ribs were orthogonally aligned with the mainstream flow, the sidewall effect was neglected to reduce computational and model training costs. Therefore, a two-dimensional, simplified cross-section (side view) was considered as the computational domain, as shown in Figure~\ref{subfig:channel-side}.

The profiled ribs, placed on the bottom and top walls in the selected region surrounding from $L/D_h$=5.0 to 5.48, seen in Figure~\ref{subfig:rectangular2D}, will be optimized with the aid of the proposed DeepONet-based surrogate. The model is trained with the pressure and Nusselt number fields data from the selected region. The profiles of the remaining ribs, located upstream and downstream from the selected region will be unchanged during the simulations. To update the profile of ribs at $L/D_h$=5.24, control points connected to each other with a spline will be used. A close-up view of the rib baseline profile with the control points is also provided in Figure~\ref{subfig:rib-points}. A summary of the channel dimensions and control points of the baseline with their initial locations is tabulated in Table~\ref{table:local-params}.

\begin{table}[!t] 
	\caption{Channel and Rib dimensions with initial control point locations.}
	\label{table:local-params}
	\begin{center}
	\begin{tabular}{lllllllllllll}
		\hline
		\multicolumn{2}{c|}{\textbf{Channel}} & \multicolumn{11}{c}{\textbf{Rib Control Points}} \\ \hline
		$W$  	& \multicolumn{1}{l|}{$H$}  	& \multicolumn{1}{l|}{Points} &1&2&3&4&5&6&7&8&9&10\\ 
		100  & \multicolumn{1}{l|}{50}  	& \multicolumn{1}{l|}{Vertical} & 0.8 & 1.6 & 2.4 & 3.2 & 3.2 & 3.2 & 3.2 & 2.4 & 1.6 & 0.8 \\ 
		$e$ = 3.2	& \multicolumn{1}{l|}{$D_h$=66.7}  	& \multicolumn{1}{l|}{Horizontal} & -1.6 & -1.6 & -1.6 & -1.6 & -0.8 & 0.8 & 1.6 & 1.6 & 1.6 & 1.6   \\
		\hline
		\multicolumn{11}{l}{* all dimensions in mm}   \\
		\hline
	\end{tabular}
	\end{center}
\end{table}

Unlike the optimization studies described in the literature, which choose control points outside of the profile and use a Bezier or NURBS-based profile update method, this paper uses many control points connected with a spline to update the rib profile. Such an approach captures geometry ranges, including sharp edges and curved profiles, while maintaining mesh quality near the updated profiles. As shown in Figure~\ref{subfig:rib-points}, ten control points are equally distributed along the rib profile. Points marked with diamond shapes (1, 2, 3, 8, 9, and 10) and blue triangles (4, 5, 6, and 7) can move in any direction on the $x$ and $y$ axes, with a maximum movement of 10\% of the initial rib height, restricted to a maximum of 0.32 mm and in the $x$ and $y$ directions, respectively. To maintain acceptable mesh distributions for all conditions, the minimum distance between two sequential points was not allowed to be lower than 5\% of the initial rib height, \ie 0.16 mm. Additionally, points marked with red cross shapes were fixed to maintain constant rib center height-to-hydraulic diameter and pitch-to-height ratios of $e/D_h$=0.048 and $P/e$=10, respectively.
\subsection{Operating Conditions and Data Reduction}
In advanced gas turbine internal cooling channels, high Reynolds numbers ranging from 10,000 to 60,000 are expected. Therefore, experimental and numerical studies ranged the Reynolds number from 5,000 to 100,000 to provide a detailed investigation of heat transfer and pressure drop behaviors in internal cooling channels. Since this paper focuses on developing novel operator learning-based surrogates for rib profile optimization, the inlet flow condition is kept constant. The inlet channel Reynolds number of 30,000 is set to satisfy the operating condition of an internal cooling channel. Additionally, due to the pressurized operating conditions, elevated coolant density is expected in these channels. Therefore, despite the small hydraulic diameter of internal cooling channels, high Reynolds numbers can still be achieved at lower coolant velocities, as shown in~\eqref{eq:Re}, resulting in a Mach (Ma) number of internal cooling channels that is mostly less than 0.3. It is worth mentioning that while performing numerical methods, the incompressible flow condition is valid and can be applied to this problem.
\begin{equation} \label{eq:Re}
	Re=\frac{\rho U_{in} D_{h}} {\mu}	
\end{equation}

To estimate the surface heat transfer coefficient in~\eqref{eq:heattc}, temperature differences between the rib-roughened surface and reference flow temperature are used in the numerical setups. It is important to note that, when comparing numerical and experimental results, the experimental measurement relies on the bulk flow temperature instead of the reference temperature used for the numerical case. The bulk flow temperature is usually calculated by linear interpolation between the inlet and outlet flow mean temperature. This paper estimates the heat transfer rate numerically based on surface and coolant reference temperatures, which are set at 350K and 300K, respectively.
\begin{equation} \label{eq:heattc}
	h = \frac{q''} {T_{s}-T_\text{bulk/ref}}
\end{equation}
Using the heat transfer coefficient ($h$), channel hydraulic diameter ($D_h$), and thermal conductivity ($k$) of the coolant (which is air in this paper), the Nusselt number -a non-dimensional function that defines the ratio of convective and conductive heat transfer rates at the boundary- can be estimated using the following equation
\begin{equation} \label{eq:No}
	Nu = \frac{h D_{h}} {k}
\end{equation}
To measure the heat transfer enhancement of the proposed rib-turbulated internal cooling channel, we can compare the channel Nusselt number to a reference Nusselt number correlation defined for fully developed smooth channels, as given in equation~\eqref{eq:Nuo}. It is important to note that because the current study uses a two-dimensional approach, the correlated Nusselt number, which was experimentally developed based on a 3D channel, must be modified by considering the reference length and channel hydraulic diameter.
\begin{equation} \label{eq:Nuo}
	Nu_o = 0.023Re^{0.8}Pr^{0.4}
\end{equation}

When measuring the hydraulic performance of a channel, the friction factor can be calculated using the pressure drop, $\Delta P$, measured between the inlet and outlet of the channel, the inlet velocity, and the channel dimensions given in equation~\eqref{eq:friction}. It is worth noting that, instead of using inlet and exit pressure values, DeepONet-based surrogates are trained using field pressure values throughout the selected region. This allows the model to provide pressure information for any given location within the selected region of the channel.

\begin{equation} \label{eq:friction}
	f = \frac{\Delta P(4D_{h}/L)} {0.5\rho U_{in}^2}
\end{equation}
To measure the pressure behavior of the channel, the friction factor was normalized using the Blasius correlation given in equation~\eqref{eq:fo} for fully developed turbulent flow in a smooth channel with a circular cross-section.
\begin{equation} \label{eq:fo}
	f_o = 0.046Re^{-0.2}
\end{equation}

{\color{black}{To understand the combined results of heat transfer enhancement and pressure drop relations in the internal cooling channel, the thermal performance using equation~\eqref{eq:TP} is evaluated.}}
\begin{equation} \label{eq:TP}
	TP = \frac{Nu/Nu_{o}} {(f/f_{o})^{1/3}}
\end{equation}

\kh{The objective of this paper is to develop a deep operator learning-based surrogate model that can map any rib profile in the selected region to the pressure and heat transfer distributions. To achieve this, data needs to be collected for a variety of rib profiles in order to use it for training. For this purpose, a cost-effective 2D numerical simulation, detailed in the following subsection, was chosen.} 
\subsection{Numerical Approach} \label{sub-sec:numerical-approach}
The numerical solutions required to train the proposed surrogates in this paper were performed using the ANSYS Workbench platform. The mesh is automatically updated for each randomly selected design point, and a total of 275 design point sets are selected. Fluent is used for numerical simulations. Since the Ma is less than 0.3, the solver settings were defined as incompressible flow based on the Reynolds Averaged Navier-Stokes (RANS) model. The generalized k-$\omega$ (GEKO) two-equations turbulence model with its default option was chosen as the turbulence model. For internal flows, the Shear Stress Transport (SST) k-$\omega$ was mainly used. However, due to the better prediction with its flexibility of tuning the parameters of the generalized k-$\omega$ model, it has also become a better option for internal flows, as detailed in \cite{2019Geko}.

The coolant properties are defined based on air at a reference temperature of 300K. Using equation~\eqref{eq:Re} and the channel hydraulic diameter, the inlet flow velocity is defined to have a Reynolds number of 30,000 as the channel's inlet boundary condition. A pressure value of 101,325 Pa is defined as the channel's outlet boundary condition. The rib-roughened walls are specified as non-slip wall boundaries with a temperature of 350K. Additionally, the top and bottom surface rib profiles are always symmetrically updated for each design point. Mesh sensitivity is applied to choose the optimum mesh size, and the results are compared to the experimental data presented in \cite{1988HAN2}. When comparing the results, attention should be paid to the channel dimensions.

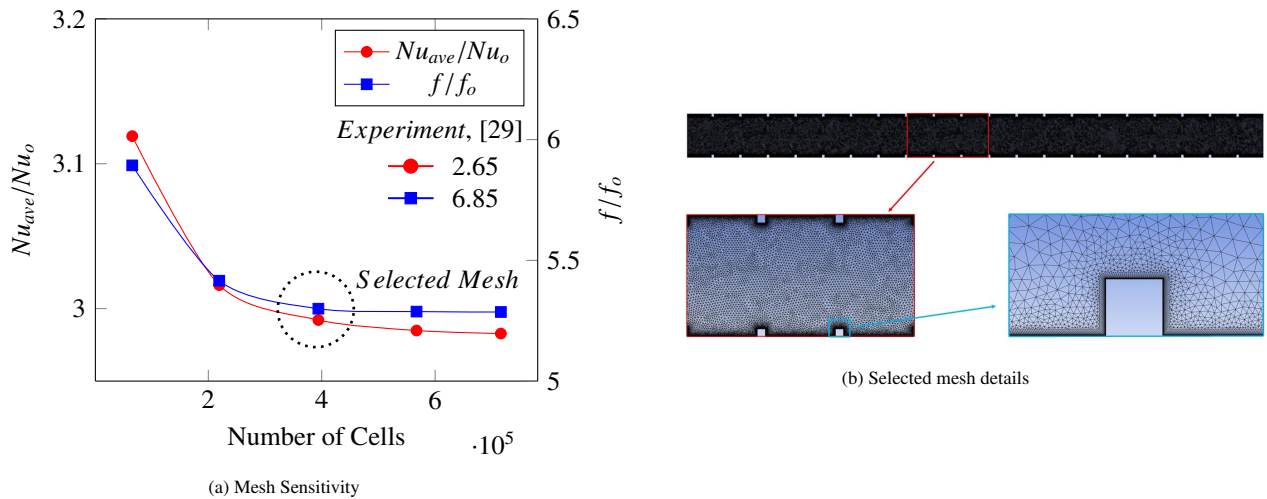
\begin{figure}[!b]
	\pgfplotsset{compat=1.3}
	\begin{subfigure}[!bl]{0.45\textwidth}
		\begin{tikzpicture}
		\begin{axis}[width=\linewidth,	
		axis y line*=left,xlabel={Number of Cells},ymin=2.95, ymax=3.2,ylabel={$Nu_{ave}/Nu_o$}]
		\addplot[smooth,mark=oplus*,red]
			coordinates{
				(65850, 3.119058277) 
				(219282, 3.016225309)
				(394112, 2.992166827)
				(567184, 2.98490254)
				(715862, 2.982809141)}; 
			\label{NusseltMesh}
		\end{axis}
		\begin{axis}[width=\linewidth,
		axis y line*=right,legend pos=north east,axis x line=none,ymin=5, ymax=6.5,ylabel={$f/f_o$}]
		\addlegendimage{/pgfplots/refstyle=NusseltMesh}\addlegendentry{$Nu_{ave}/Nu_o$}
		\addplot[smooth,mark=square*,blue]
			coordinates{
				(65850, 5.89336253) 
				(219282, 5.415727957)
				(394112, 5.29978334)
				(567184, 5.287817721)
				(715862, 5.286145211)};
			\label{PressureMesh}
		\addlegendimage{/pgfplots/refstyle=PressureMesh}\addlegendentry{$f/f_o$}
		\end{axis}
		\draw [line width=1pt,dotted] (2.9,0.95) circle (5mm);
		\node at (4.5, 1.35) {$Selected$ $Mesh$};
		\node at (4.40, 3.3) {$Experiment$, \cite{1988HAN2}};
		\fill[red]	(4.15,2.85)	circle (1mm); \node at (5.0, 2.85) {$2.65$};
		\draw [line width=0.75pt, red] (3.85,2.85) -- (4.45, 2.85); 
		\filldraw[draw=blue,fill=blue] (4.075,2.35) rectangle (4.225,2.50);
		\node at (5.0, 2.425) {$6.85$};
		\draw [line width=0.75pt, blue] (3.85,2.425) -- (4.45, 2.425);
		\end{tikzpicture}
		\caption{Mesh Sensitivity}
		\label{sub-fig:MeshSens1}	
	\end{subfigure}
	\hspace*{40pt}
	\begin{subfigure}[br]{0.4\textwidth}
		\centering{\includegraphics[width=3.0in]{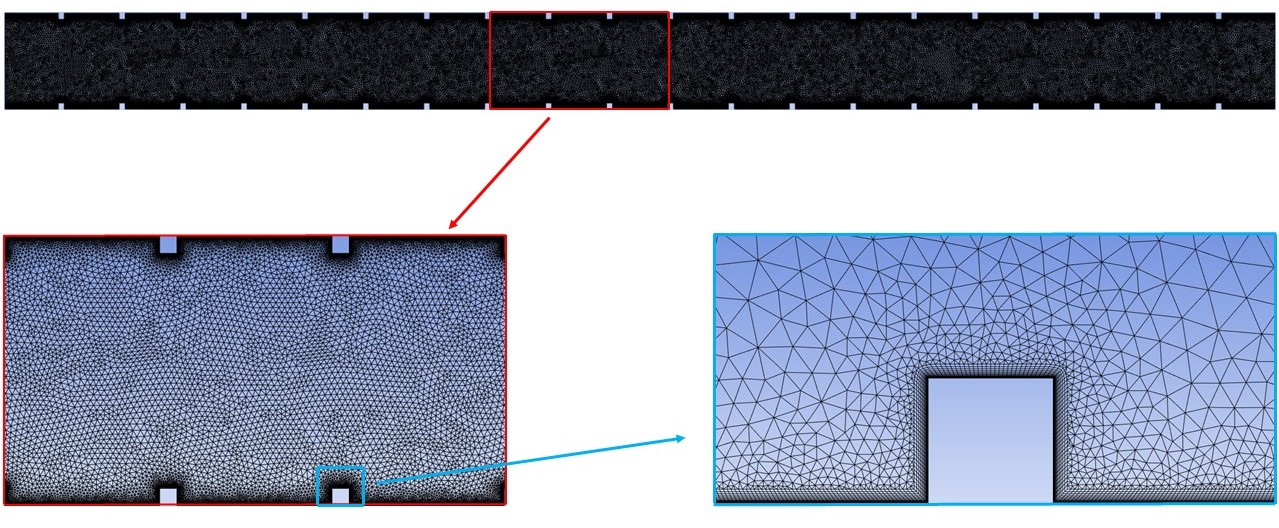}}
		\caption{Selected mesh details}
		\label{sub-fig:Mesh-details}	
	\end{subfigure}
	\caption{Schematic of orthogonal rib roughened internal cooling channel}
	\label{MeshSens}
\end{figure}

\subsubsection{Grid sensitivity and Experimental validation}
To ensure the accuracy of the numerical approach and results, mesh sensitivity analysis was performed for a 2-dimensional internal cooling channel with square ribs. First, the first-layer mesh size was predicted for $y^{+}$ values less than one, considering the channel's bulk flow velocity. Wall element sizes were calculated based on the first layer's grid aspect ratio, and the overall mesh element size was then defined. Several meshes with different intensities were created, and simulations were performed for each mesh until all residuals were less than 10$^{-6}$.

Figure~\ref{sub-fig:MeshSens1} shows the channel average Nusselt number and friction factor ratios for each mesh. The mesh was chosen when the error percentage between two sequential meshes became lower than 1$\%$ and 2$\%$ for Nusselt number and friction factor ratios, respectively. As depicted in Figure~\ref{sub-fig:MeshSens1}, the mesh with 394,112 cells was selected for further investigation and data production, considering the computational cost and a large number of design point sets. Figure~\ref{sub-fig:Mesh-details} provides a close-up view of the mesh. Channel averaged experimental and numerical results were also provided in Figure~\ref{sub-fig:MeshSens1} for comparison. It shows that the Nusselt number of the 2D numerical solution is almost 13$\%$ higher, while the friction factor is almost 22$\%$ lower than that of the experimental results. The high Nusselt number in the 2D cases is due to the absence of sidewalls and their heating effects on the bulk flow temperature. Similarly, the 2D numerical case friction factor is lower than the experimental case due to the absence of the side wall shear stresses.

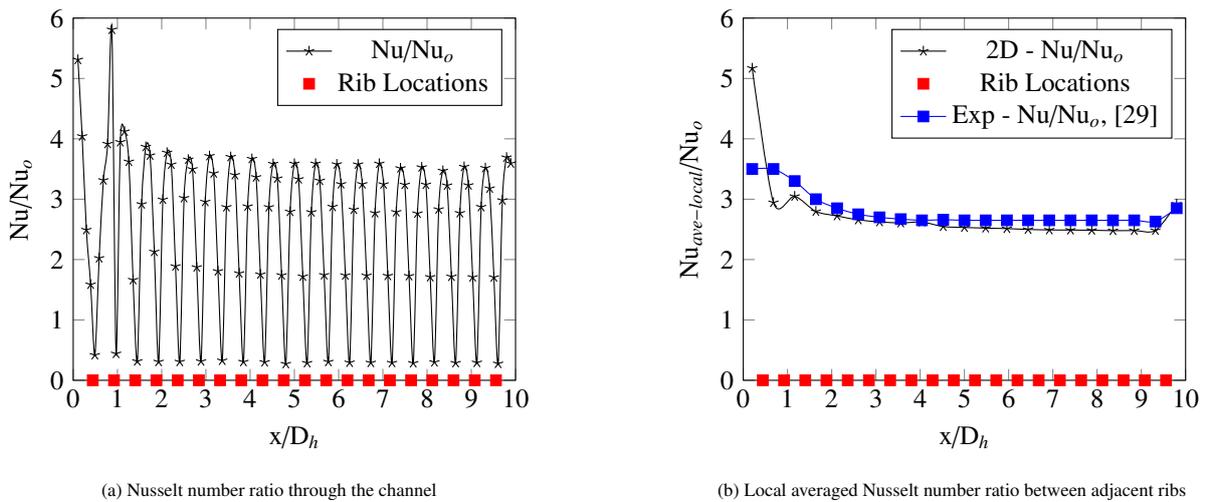
\begin{figure}[!b] 
	\begin{subfigure}[!bl]{0.45\textwidth}
	\centering
	\begin{tikzpicture}
	\begin{axis}[width=\linewidth,
		xmin = 0,xmax=10,	xlabel={x/D$_h$},ymin=0, ymax=6, ylabel={Nu/Nu$_o$},	legend pos=north east, xtick={0,1,...,10}, ytick={0,1,2,...,6},]
		\addplot [smooth, black, mark=star] 
		table{data/Nuupdate.txt}; 	\label{Nusselt}; \addlegendentry{Nu/Nu$_o$}
		\addplot+ [only marks, mark=square*, mark size=2pt,mark options={draw=red,fill=red}]
		table{data/Rib.txt};	\label{RibLocation}; \addlegendentry{Rib Locations}	
	\end{axis}
	\end{tikzpicture}
	\caption{Nusselt number ratio through the channel}
	\label{sub-fig:Nuratio2}
	\end{subfigure}
\hspace{40pt}
	\begin{subfigure}[!br]{0.45\textwidth}
	\begin{tikzpicture}
		\begin{axis}[width=\linewidth,
			xmin = 0.0,xmax=10,	xlabel={x/D$_h$},	ymin=0, ymax=6, ylabel={Nu$_{ave-local}$/Nu$_o$},	legend pos=north east,xtick={0,1,...,10}, ytick={0,1,2,...,6},]
			\addplot [smooth, black, mark=star] 
			table{data/Nuwithoutrib.txt}; 	\label{Nusseltlocal}; \addlegendentry{2D - Nu/Nu$_o$}
			\addplot+ [only marks, mark=square*, mark size=2pt,mark options={draw=red,fill=red}]
			table{data/Rib.txt};	\label{RibLocation1}; \addlegendentry{Rib Locations}	
			\addplot [smooth, blue, mark=square*] 
			table{data/Nuwithoutrib1.txt}; 	\label{Nusseltex}; \addlegendentry{Exp - Nu/Nu$_o$, \cite{1988HAN2}}
		\end{axis}
	\end{tikzpicture}
	\caption{Local averaged Nusselt number ratio between adjacent ribs}
	\label{sub-fig:Nuratio3}
\end{subfigure}
\caption{Nusselt number ratios}
\label{Nuratio1}
\end{figure}

Additionally, to illustrate the heat transfer behavior, the distribution of the Nusselt number ratio in the square rib-roughened surface of the internal cooling channel is presented in Figure~\ref{sub-fig:Nuratio2}. The average Nusselt number ratios between each adjacent rib are displayed and compared with the experimental data in Figure~\ref{sub-fig:Nuratio3}. It is important to note that the referenced experimental study \cite{1988HAN2} does not measure rib surface heat transfer results, but instead provides the Nusselt number ratios between each adjacent rib. Figure~\ref{sub-fig:Nuratio3} shows that the averaged Nusselt number ratios for both the experimental and numerical solutions are almost identical, except in the entrance region due to the entrance effect. As mentioned earlier, to compare the experimental and 2D numerical solutions, it is necessary to convert the Nusselt number correlation given in equation~\eqref{eq:Nuo} from 3D to 2D conditions by dividing it by the channel hydraulic diameter and multiplying it by the reference length used for the 2D simulation setup. {\color{black}{With the validated numerical approach presented in this section, we will construct the dataset to train the proposed DeepONet-based surrogate model in the following section.}}

\section{Deep Operator Network-based Surrogate Models} \label{sec:DeepONet}
{\color{black}{This section focuses on developing surrogate models for the operator mapping between control points of the rib profile $u \in \mathcal{U}$ and the distribution of pressure $P(u,y)$ and Nusselt number $Nu(u,y)$. Where $\mathcal{U}$ and $Y$ are the space of all rib profiles and the domain of all streamwise coordinates (i.e., the output function location), respectively. Also, $y$ is any streamwise channel \kh{location in the $Y$ domain.} Formally, we define the space of all rib profiles as the set of all control points whose distance from the original square rib profile (denoted as~$u_o$) does not exceed $0.32$ mm, \ie $\mathcal{U}:=\{u \in \mathbb{R}^m~:~\|u-u_o\| < 0.32\text{ mm} \}$. Where $m$ is the total number of vertical and horizontal locations of control points on each rib profile, the current model has ten control points, and each of the points can be defined in two-dimensional vector space in the vertical and horizontal directions that yields $m$ to be 20 for any rib profile.

For locally averaged pressure through the selected regions, the domain~$Y$ of all streamwise directions corresponds to the uniform partition of the interval $[L/D_h=5,5.48]$  using $n_\text{loc}=300$ points. On the other hand, since the Nu number is measured on the surface for a selected streamwise location, several Nu numbers can be seen, especially on the vertical surface of the rib, which causes misleading the model during the training. Therefore, instead of streamwise location, total surface length, including the rib surface and upstream and downstream surfaces of the selected region, is divided into uniform $n_\text{loc}=300$ points to define the domain $Y$. }}

\begin{figure}[!b] 
	\centering
		\begin{tikzpicture}	
		\draw [->, line width=1pt] (0.25,1.25)-- (1.0,1.25);
		\node[text width=5.0,fill=white] at (0.4, 1.5) {\textbf{u}};
		\draw [line width=1pt] (1.25,0) -- (1.25,2.5) -- (2.25,2.5) -- (2.25,0) -- (1.25,0);
		\node[text width=5.0,fill=white] at (1.7, 2.10) {$u_1$};
		\node[text width=5.0,fill=white] at (1.7, 1.6) {$u_2$};
		\node[text width=5.0,fill=white] at (1.7, 1.1) {$u_3$};
		\draw [line width=1pt, dotted] (1.6,0.7)-- (1.9,0.7);
		\node[text width=5.0,fill=white] at (1.7, 0.3) {$u_m$};
		\node[text width=5.0,fill=white] at (1.35, -0.33) {(input)};
		\node[] at (1.8, -0.80) {(Control design points)};
		\draw [->, line width=1pt] (2.5,1.25)-- (3.25,1.25);
		\filldraw[draw=black,fill=teal] (3.5,0.9) rectangle (5.75,1.6);
		\node[text width=50] at (4.65, 1.25) {Branch Net};
		\draw [->, line width=1pt] (6,1.25)-- (7.2,1.95);
		\draw [->, line width=1pt] (6,1.25)-- (7.2,1.55);
		\draw [->, line width=1pt] (6,1.25)-- (7.2,0.5);
		\draw [line width=1pt] (7,0) -- (7,2.5) -- (8,2.5) -- (8,0) -- (7,0);
		\node[text width=5.0,fill=white] at (7.45, 2.10) {$b_1$};
		\node[text width=5.0,fill=white] at (7.45, 1.6) {$b_2$};
		\node[text width=5.0,fill=white] at (7.45, 1.1) {$b_3$};
		\draw [line width=1pt, dotted] (7.4,0.7)-- (7.7,0.7);
		\node[text width=5.0,fill=white] at (7.45, 0.32) {$b_p$};
		\draw [->, line width=1pt] (2.5,-2.25)-- (3.25,-2.25);
		\node[text width=5.0,fill=white] at (2.75, -1.95) {\textbf{y}};
		\node[text width=5.0,fill=white] at (2.1, -2.65) {(location)};
		\filldraw[draw=black,fill=teal] (3.5,-1.9) rectangle (5.75,-2.6);
		\node[text width=50] at (4.65, -2.25) {Trunk Net};
		\draw [->, line width=1pt] (6,-2.25)-- (7.2,-2.95);
		\draw [->, line width=1pt] (6,-2.25)-- (7.2,-1.9);
		\draw [->, line width=1pt] (6,-2.25)-- (7.2,-1.5);
		\draw [line width=1pt] (7,-1) -- (7,-3.5) -- (8,-3.5) -- (8,-1) -- (7,-1);
		\node[text width=5.0,fill=white] at (7.45, -1.35) {$t_1$};		
		\node[text width=5.0,fill=white] at (7.45, -1.85) {$t_2$};
		\node[text width=5.0,fill=white] at (7.45, -2.35) {$t_3$};
		\draw [line width=1pt, dotted] (7.4,-2.75)-- (7.7,-2.75);
		\node[text width=5.0,fill=white] at (7.45, -3.10) {$t_p$};
		\draw [->, line width=1pt] (8.0,-2.25)-- (9.25,-0.75);
		\draw [->, line width=1pt] (8.0,1.2)-- (9.25,-0.25);	
		\draw [line width=1pt,fill=lightgray] (9.5,-0.5) circle (3mm);
		\node[text width=5] at (9.405, -0.5) {\LARGE $\times$};
		\draw [->, line width=1pt] (9.9,-0.5)-- (10.5,-0.5);
		\node[text width=5.0] at (10.7, -0.525) {$G(u)(y)$};
		\end{tikzpicture}			
	\caption{The Deep Operator Network~(DeepONet) architecture. The Branch Net processes the control design points and the Trunk Net the $x$-wall location. The prediction is obtained using a dot product between the Branch and Trunk Nets' outputs.}
	\label{fig:DeepONet}
\end{figure}
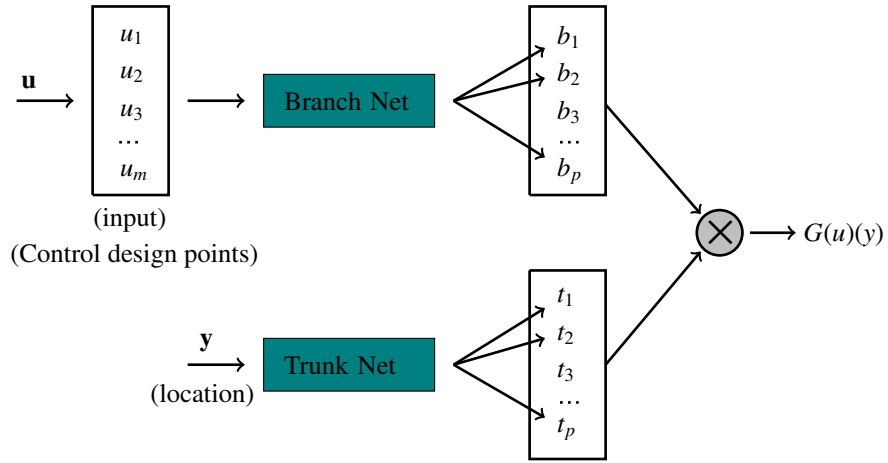

We use Deep Operator Networks (DeepONet)~\cite{2021Lulu1} to approximate the aforementioned operator mapping. DeepONet, denoted as $G^\alpha_{\theta^\alpha}$ where $\alpha \in \{P,Nu\}$, is based on the universal approximation theorem of nonlinear operators~\cite{1995Chen1} and can approximate any nonlinear continuous operator (i.e., a mapping between infinite-dimensional spaces). DeepONet uses a trainable linear representation (with trainable parameters $\theta^\alpha$), as illustrated in Figure~\ref{fig:DeepONet}, to approximate the pressure and Nusselt number distributions as follows:
\begin{align*}
   P(u,y) &\approx G^P_{\theta^P}(u)(y) = \sum_{k=1}^p b^{P}_k(u) \cdot t^{P}_k(y), \\
   Nu(u,y) &\approx G^{Nu}_{\theta^{Nu}}(u)(y) = \sum_{k=1}^p b^{Nu}_k(u) \cdot t^{Nu}_k(y). 
\end{align*}

To implement the trainable linear representation shown above, we need to construct and train two sub-neural networks: the Branch network $b^\alpha$ and the Trunk network $t^\alpha$. For the sake of simplicity, we will omit the superscript $\alpha$, since designing DeepONets for the pressure and Nusselt number surrogates requires similar steps. More specifically, the Branch network processes the control design points $u \in \mathcal{U}$ and outputs a vector of trainable coefficients $b(u) = (b_1, \ldots, b_p)$. Similarly, the Trunk network takes as input a given output function location $y \in Y$ and outputs a vector of trainable basis functions $t(y) = (t_1, \ldots, t_p)$. By combining the coefficients with the basis functions using a dot product, we obtain the DeepONet linear representation $G_\theta(u)(y)$.

To train and implement the DeepONets, we optimize the following loss function:
\kh{$$\mathcal{L}(\theta^\alpha;\mathcal{D}^\alpha) = \sum_{j=1}^N \|\alpha_j(u_j,y_j) - G^\alpha_{\theta^\alpha}(u_j)(y_j)\|^2$$}
where $\alpha \in \{P,Nu\}$ and using the dataset of $N$ i.i.d. triplets:
$$\mathcal{D}^P =\{(u_j, y_j), P_j\}_{j=1}^N \quad \text{ and/or  } \quad \mathcal{D}^{Nu} =\{(u_j, y_j), Nu_j\}_{j=1}^N.$$
In the above, for simplicity, we have assumed that for each rib design input $u_j$, we sample one \kh{channel location} $y_j \in Y$. In practice, however, for each $u_j$, we can sample a set of $q$ \kh{channel location} inputs $\{y_j^{(i)}\}_{i=1}^q$. In our numerical experiments, for each~$u_j$, we used $q=320$  sampled locations during training.
\subsection{Training and Testing the DeepONet-based Surrogate Model}
This section presents the training and testing protocols used to design the proposed DeepONet-based surrogate models. To begin, let us summarize the validated numerical approach (detailed in Section~\ref{sec:problem-formulation}) used to generate the training and testing datasets.

\textit{Dataset generation.} To train and test the DeepONet-based surrogate models in this paper, 2-dimensional numerical simulations were performed for each rib profile of control points $u_j \in \mathcal{U}$, for $j=1,\ldots,N$. Each rib profile includes vertical and horizontal location information for all control points. A spline connects the 10 control points to reshape the rib profile. The initial location information or reference control points set is shown in Figure~\ref{subfig:rib-points}. As mentioned in the previous section, the rib profile, located at $L/D_h=5.24$ on the top and bottom walls, was symmetrically updated for each case.

To obtain training targets for the DeepONet surrogate models, Nusselt number ratio and pressure results were collected at a distance of $P/e=5$ from the upstream and downstream of the center of the selected rib. Specifically, the training data was collected between $L/D_h=5.0$ and $L/D_h=5.48$, as described by the domain $Y$ and shown in Figure~\ref{subfig:rectangular2D}.

A total of $N=275$ rib profiles of control points were randomly selected for training and testing the DeepONet model, with 90\% used for training and 10\% for testing. Unlike other deep learning-based approaches in the literature that use algorithms or uniform sampling methods to create the training and testing datasets, this paper used a simple random selection. This means that the sequential data control points became independent, resulting in a more reliable surrogate model that can be generalized.

\textit{Neural networks.} We constructed the Branch and Trunk networks using feed-forward neural networks (FNNs). To determine the best FNN architectures, we used a simple hyper-parameter optimization routine. This resulted in Branch and Trunk networks with five hidden layers, each containing 500 neurons. The DeepONet model was trained using the Adam optimizer with default hyperparameters and an initial learning rate of $\eta = 5 \times 10^{-3}$. To enhance the training process, we utilized the ``reduce on plateau'' learning rate scheduler. This scheduler automatically adjusts the learning rate when the loss function reaches a plateau or starts to increase. This helps prevent the model from becoming trapped in a local minimum and helps achieve faster convergence.
\subsection{DeepONet-based Surrogate Models Numerical Results}
This section presents a series of results that showcase the prediction and generalization capacity of the proposed DeepONet-based surrogate models.

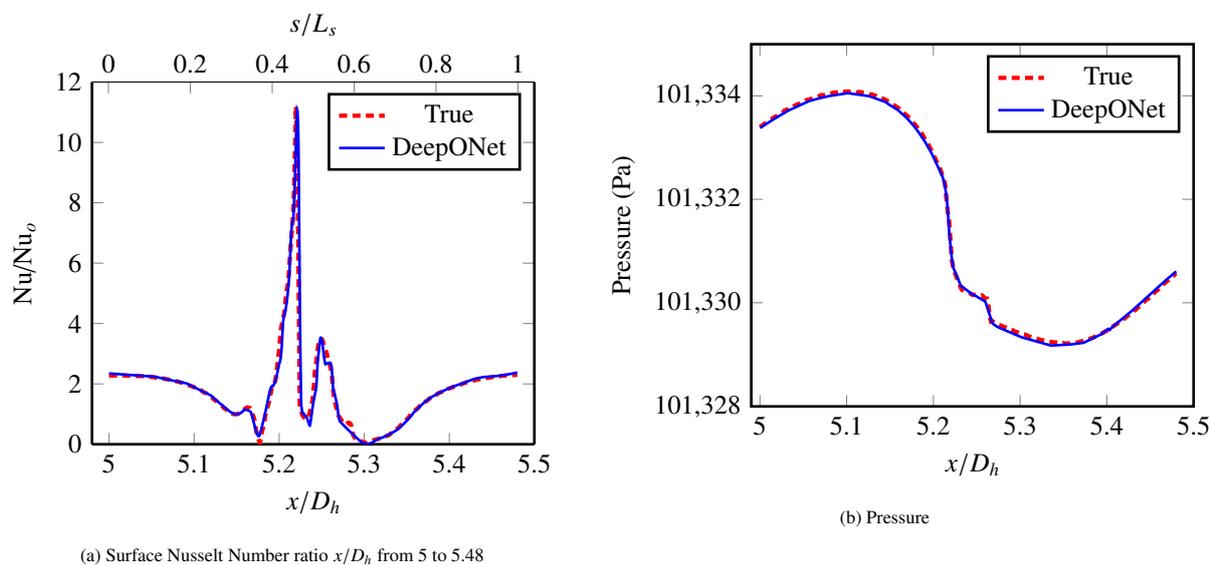
\begin{figure}[!b] 
	\begin{subfigure}[!bl]{0.45\textwidth}
	\centering
	\begin{tikzpicture}
	\begin{axis}[width=\linewidth, xlabel={$x/D_h$}, xmin = 4.98, xmax=5.5, ymin=0, ymax=12, ylabel={Nu/Nu$_o$}, legend pos=north east, line width=1pt, xtick={5,5.1,...,5.5}, axis x line*=bottom, ytick={0,2,...,12},]
		\addplot [densely dashed, red, ultra thick] 
		table{data/Nudata.txt}; 	\label{Nusselty}; \addlegendentry{True}
        \addplot [smooth, blue]
        table{data/NuPredic.txt};	\label{RibLocationy}; \addlegendentry{DeepONet}	
	\end{axis}
    \begin{axis}[width=\linewidth, xmin = -0.04, xmax=1.04, ymin=-2, ymax=12, xtick={0,0.2,...,1}, xlabel={$s/L_s$}, axis x line*=top, axis y line=none,]\end{axis}
	\end{tikzpicture}
	\caption{Surface Nusselt Number ratio $x/D_h$ from 5 to 5.48}
	\label{subfig:prediction-Nu}
	\end{subfigure}
    \hspace{10pt}
	\begin{subfigure}[!br]{0.45\textwidth}
    \centering
	\begin{tikzpicture}
		\begin{axis}[width=\linewidth, xmin = 4.99, xmax=5.5, xlabel={$x/D_h$}, ymin=101328, ymax=101335, ylabel={Pressure (Pa)}, legend pos=north east, xtick={5,5.1,...,5.5},line width=1pt, yticklabel style={scaled ticks=false, /pgf/number format/fixed, /pgf/number format/precision=1}]           
            \addplot [densely dashed, red, ultra thick] 
			table{data/Prdata.txt};	\label{RibLocationz}; \addlegendentry{True}	
			\addplot [smooth, blue] 
			table{data/PrPredic.txt}; 	\label{Nusseltez}; \addlegendentry{DeepONet}
		\end{axis}
	\end{tikzpicture}
	\caption{Pressure}
	\label{subfig:prediction-P}
\end{subfigure}
\caption{Comparison of the true distribution, obtained using the validated numerical approach, and the prediction of DeepONet (i) for a rib profile $u_j \in \mathcal{U}$ selected at random from the testing dataset, and (ii) around the profiled ribs, i.e., over a uniform partition of $n_\text{loc}=300$ locations within the output domain $Y$.}
\label{NuandPrPredic}
\end{figure}

{\color{black}{Figures~\ref{subfig:prediction-Nu} and \ref{subfig:prediction-P} show the distribution of Nusselt number ratio and pressure}} (i) for a rib profile $u_j \in \mathcal{U}$ selected at random from the testing dataset, and (ii) around the profiled ribs, i.e., over a uniform partition of $n_\text{loc}=300$ locations within the output domain $Y$. The results illustrate excellent agreement between the true distribution, obtained using the validated numerical approach, and the prediction of DeepONet. {\color{black}{It is worth mentioning that since the trunk of DeepONet is constructed as a one-dimensional location space in this study, as illustrated in Figure~\ref{fig:DeepONet},  for the Nusselt number surrogate model, location input is considered as the surface length from $x/D_h$ = 5 to 5.48 shown in Figure~\ref{subfig:prediction-Nu}. Thus, the two-dimensional vertical and horizontal locations of the rib profile are reduced to the one-dimensional surface domain s ranging from 0 to $L_s$.}}

\kh{To evaluate the efficiency of the proposed DeepONet-based surrogates, we calculated the mean and standard deviation of the simulation time required to obtain the Nusselt number and pressure distributions for $30$ random test rib profiles using both the validated numerical approach and the DeepONet-based surrogates. Tables \ref{table:efficiency-Nu} and \ref{table:efficiency-P} show that the DeepONet-based surrogates significantly reduce computational costs.}
\begin{table}[t!]
\caption{\kh{The mean and standard deviation of the simulation time (in seconds) required to obtain the Nusselt number ($Nu$) distribution for $30$ random test rib profiles using both the validated numerical approach and the DeepONet-based surrogates.}}
\centering
\begin{tabular}{c c c} 
 \hline
  & \textbf{Numerical Approach} & \textbf{DeepONet} \\ [0.5ex] 
 \hline
 \textbf{mean} (s) & $7.21 \cdot 10^2$ & $5.39 \cdot 10^{-4}$ \\ 
 \textbf{st.dev.} (s) & $1.09 \cdot 10^1$  & $1.67 \cdot 10^{-5}$\\ [1ex] 
 \hline
\end{tabular}
\label{table:efficiency-Nu}
\end{table}
\begin{table}[t!]
\caption{\kh{The mean and standard deviation of the simulation time (in seconds) required to obtain the pressure ($P$) distribution for $30$ random test rib profiles using both the validated numerical approach and the DeepONet-based surrogates.}}
\centering
\begin{tabular}{c c c} 
 \hline
  & \textbf{Numerical Approach} & \textbf{DeepONet} \\ [0.5ex] 
 \hline
 \textbf{mean} (s) & $7.21 \cdot 10^2$& $9.39 \cdot 10^{-4}$ \\ 
 \textbf{st.dev.} (s) & 1.09 $\cdot 10^1$ & 1.18 $\cdot 10^{-5}$\\ [1ex] 
 \hline
\end{tabular}
\label{table:efficiency-P}
\end{table}

Finally, to test the generalization capacity of the proposed DeepONet-based surrogate models, we computed the mean and standard deviation of the $L_1-$relative error (see~\eqref{eq:L1-error}, where $\|\cdot\|_1$ is the $L_1$ norm, $\omega$ the true trajectory, and $\hat{\omega}$ the DeepONet predicted trajectory) between the true and predicted distributions of pressure and Nusselt number ratio for all $N_\text{test}= { \color{black} 27}$ rib profiles $u_j \in \mathcal{U}$. These statistics (depicted in Table~\ref{table:DeepONet-results}) reveal that the proposed surrogates can keep the errors below 8\% for both distributions, even with a relatively small dataset. We note that these statistics can be improved by augmenting the training dataset, although acquiring new data can be costly in practice. To better handle this scenario,  in the next section, we propose a Bayesian framework for quantifying the uncertainty of DeepONet surrogate models.
\begin{table}[t!]
\caption{The mean and standard deviation of the $L_1$-relative error  between the true and DeepONet predicted distributions of pressure and Nusselt number around the profiled ribs.}
\centering
\begin{tabular}{c c c} 
 \hline
  & \textbf{Pressure} ($P$) & \textbf{Nusselt Number} ($Nu$) \\ [0.5ex] 
 \hline
 \textbf{mean}~$L_1$ & 5.53\% & 7.50\% \\ 
 \textbf{st.dev.}~$L_1$ & 6.67\%  & 1.67\%\\ [1ex] 
 \hline
\end{tabular}
\label{table:DeepONet-results}
\end{table}

\begin{align} \label{eq:L1-error}
L_{1}\text{error \%} =  100 \% \cdot \frac{ \|\omega - \hat{\omega}\|_1}{\|\omega \|_1}.
\end{align}
\section{Quantifying the Uncertainty of DeepONet-based Surrogates} \label{sec:B-DeepONet}
We recognize that obtaining enough data to train surrogate DeepONet models can be computationally challenging in practice. As a result, a rib optimization design routine that uses a DeepONet-based surrogate trained with limited data may produce adversarial designs - fake optimal designs that could potentially compromise the operation of internal cooling channels. To address this issue, in this section, we develop a Bayesian framework for quantifying the uncertainty of DeepONet models, \ie Bayesian DeepONet (B-DeepONet)~\cite{lin2023b} surrogate models for the Nusselt number and pressure distribution in this section. These surrogates allow for the quantification of the epistemic uncertainty that arises from training the DeepONet parameters $\theta$ using limited data. By quantifying the uncertainty, B-DeepONet surrogates protect the rib optimization process by (i) identifying and eliminating designs that are likely to be adversarial and (ii) selecting designs that are likely correct but suboptimal. We present the details for designing B-DeepONet next.

\subsection{Bayesian DeepONet Design} \label{subsec:B-DeepONet-design}
The conventional optimization framework used to train DeepONets does not accurately quantify uncertainty, which is crucial for creating credible intervals for Nusselt number and pressure predictions. This lack of credible intervals leads to unreliable DeepONet predictions and raises doubts about the reliability of DeepONet-based optimized rib designs. However, quantifying the uncertainty associated with limited training data and neural network over-parametrization is a challenging task. And this challenge is even greater in operator learning, as it involves mappings between infinite-dimensional spaces.

In this paper, we address this challenge by developing a Bayesian DeepONet (B-DeepONet)~\cite{lin2023b}. B-DeepONet can create estimators and credible intervals for the operator that maps a given rib profile $u \in \mathcal{U}$ to the Nusselt number $Nu(u,y)$ or pressure $P(u,y)$ distributions for any given $y \in Y$.

In this Bayesian DeepONet framework, which we first proposed in~\cite{lin2023b,moya2023deeponet}, our goal, given the training dataset $\mathcal{D}$, is to construct a distribution $p(G|(u,y),\mathcal{D})$ that can predict the operator value of $G$ (either Nusselt number $Nu$ or pressure $P$) based on the input rib profile $u$ and at any new $x$-wall location $y$. To achieve this, we first assume the following factorized Gaussian likelihood function for the data:
\begin{gather} \label{eq:gaussian-likelihood}
p(G|(u,y),\theta) = \mathcal{N}(G|G_\theta(u)(y), \text{diag}(\Sigma^2)) = \prod_{j=1}^N \mathcal{N}(G_j|G_\theta(u_j)(y_j), \sigma),
\end{gather}
where the output $G_\theta(u)(y)$ is the mean of the Gaussian distribution assumed for $G$, given the rib profile $u$ at location $y$, and $\text{diag}(\Sigma^2)$ is a diagonal covariance matrix with $\Sigma^2 = (\sigma^2,\ldots,\sigma^2)$ on the diagonal~\cite{psaros2023uncertainty}. Note that~$\sigma$ can be assumed or estimated from the data.

The Nusselt number or pressure distribution $G$ for a rib profile $u$ at a location $y$, given the training data $\mathcal{D}$, is the random variable $(G|(u,y),\mathcal{D})$. To obtain the density of this random variable, we need to integrate out the model parameters as follows:
$$p(G|(u,y), \mathcal{D}) = \int p(G|(u,y),\theta)p(\theta | \mathcal{D})d\theta.$$
Here, $p(\theta | \mathcal{D})$ represents the posterior distribution of the trainable parameters. This distribution enables us to quantify the \textit{epistemic uncertainty}, which refers to the uncertainty related to the trainable parameters $\theta$~\cite{psaros2023uncertainty,lin2023b}.

To obtain this posterior, we use Bayes' rule:
$$p(\theta|\mathcal{D}) \propto p(\mathcal{D}|\theta) p(\theta),$$
where $p(\theta)$ is the \textit{prior} distribution of the parameters and $p(\mathcal{D}|\theta)$ is the \textit{data likelihood}, i.e., $p(\mathcal{D}|\theta) = \prod_{j=1}^N p(G_j|(u_j, y_j), \theta)$, which we calculate using the DeepONet forward pass and the i.i.d training dataset~$\mathcal{D}$.

Acquiring the posterior distribution using Bayes' rule is often computationally and analytically intractable~\cite{psaros2023uncertainty}. Therefore, in our previous work~\cite{lin2023b}, we approximated the posterior distribution through samples obtained from it. Specifically, we obtained an $M$-ensemble of $\theta$ samples, denoted as $\{\theta_k\}_{k=1}^M$, as described below.
\subsection{Sampling the $M$-ensemble $\{\theta_k\}_{k=1}^M$}  \label{subsec:sampling-M-ensemble}
To obtain the $M$-ensemble of parameters $\{\theta_k\}_{k=1}^M$, B-DeepONet uses the stochastic gradient replica exchange Langevin diffusion (SG-reLD), which we developed and studied in~\cite{lin2023b,deng2020non}. As demonstrated in our previous work, SG-reLD enjoys theoretical guarantees beyond convex scenarios, effectively handles large datasets, and accelerates convergence to the posterior distribution $p(\theta|\mathcal{D})$.

Specifically, SG-reLD uses two Langevin diffusions to describe the stochastic dynamics of $\theta$, along with a stochastic process that allows the diffusions to swap simultaneously. The high-temperature diffusion enables exploration of the parameter space, facilitating convergence to the flattened distribution of $\theta$. The low-temperature diffusion exploits the same parameter space, enabling faster convergence to local minima $\theta^*$. By swapping the diffusions, SG-reLD effectively escapes local minima and allows $\theta_k$ to converge faster to the desired posterior $p(\theta|\mathcal{D})$. For more details about employing SG-reLD with B-DeepONets, please refer to our previous paper \cite{lin2023b}.

In practice, we can use the $M$-ensemble $\{\theta_k\}_{k=1}^M$ obtained using SG-reLD to fit a parametric distribution, such as the Gaussian distribution $\mathcal{N}(\bar{\mu}(u)(Y_{\text{test}}), \bar{\sigma}_e(u)(Y_{\text{test}}))$ for an arbitrary mesh~$Y_{\text{test}}$ of $x$-wall locations (in our experiments, $Y_\text{test}$ is a uniform partition of size $n_\text{loc}=300$ within $Y$). The parameters of this distribution are given by:
\begin{align*}
    \bar{\mu}(u)(Y_{\text{test}}) &= \frac{1}{M} \sum_{k=1}^M G_{\theta_k}(u)(Y_{\text{test}}), \\
    \bar{\sigma}_e(u)(Y_{\text{test}}) &= \frac{1}{M}\sum_{k=1}^M \left(G_{\theta_k}(u)(Y_{\text{test}}) - \bar{\mu}(u)(Y_{\text{test}})  \right).
\end{align*}
Sampling from the aforementioned distribution allows for estimating credible sets for Nusselt number and pressure predictions. This also enables sampling of more reliable predictions, which can help prevent adversarial rib designs as demonstrated in Section~\ref{sec:optimization}.
\subsection{B-DeepONet-based Surrogate Model Numerical Results} \label{subsec:B-DeepONet-results}
This section briefly showcases how our Bayesian DeepONet framework constructs credible/confidence intervals (CI) for test trajectories that are not included in the training dataset $\mathcal{D}$. To make the problem more realistic, we trained the B-DeepONet with $q=100$ sampled locations, which is one-third of what we used to train the deterministic DeepONet.

Figures~\ref{fig:B-DeepONet-Nusselt} and \ref{fig:B-DeepONet-pressure} show the confidence intervals we constructed for the distributions of the Nusselt number ratio and pressure, respectively. These confidence intervals (CIs) are computed (i) for a rib profile $u_j \in \mathcal{U}$ selected at random from the testing dataset, and (ii) around the profiled ribs, i.e., over a uniform partition of $n_\text{loc}=300$ locations within the output domain $Y$. These results illustrate that the confidence intervals (CIs) can capture the entire true distribution obtained using the validated numerical approach. It is worth noting that for the optimization described in the next section, we sample predictions from the fitted Gaussian distribution $\mathcal{N}(\bar{\mu}(u)(Y_{\text{test}}), \bar{\sigma}e(u)(Y_{\text{test}}))$, instead of computing confidence intervals. Sampled predictions are more reliable and will help prevent adversarial designs.

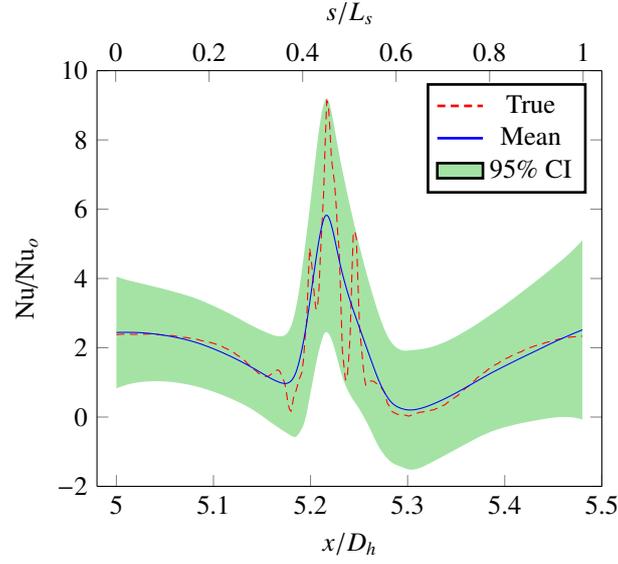
\begin{figure}[!t]
  \centering
  \begin{tikzpicture}
        \begin{axis}[width=\linewidth*0.5, xlabel={$x/D_h$}, xmin = 4.98, xmax=5.5, ymin=-2, ymax=10, ylabel={Nu/Nu$_o$}, legend pos=north east, line width=1pt, yticklabel style={scaled ticks=false}, axis x line*=bottom]    
             \addplot[name path=Nu_true,densely dashed, color=red] table{data/Nu_true.txt};
             \addlegendentry{True}
             \addplot[ name path=Nu_mean, smooth, color=blue] table{data/Nu_mean.txt};
             \addlegendentry{Mean}
             \addplot[name path=Nu_lower,  fill=none, draw=none,forget plot] table{data/Nu_lower.txt};
             \addplot[name path=Nu_upper, fill=none, draw=none,forget plot] table{data/Nu_upper.txt};
            \definecolor{mycolor}{RGB}{75,200,75}
            \addplot[fill=mycolor, fill opacity=0.5] fill between[of=Nu_upper and Nu_lower, ];
            \addlegendentry{$95 \%$ CI}
        \end{axis}
        \begin{axis}[width=\linewidth*0.5, xmin = -0.04, xmax=1.04, ymin=-2, ymax=10, xtick={0,0.2,...,1}, xlabel={$s/L_s$}, axis x line*=top, axis y line=none,]\end{axis}
  \end{tikzpicture}
  \caption{B-DeepONet mean prediction and confidence interval (CI) for Nusselt number ratio true distribution, obtained using the validated numerical approach (i) for a rib profile $u_j \in \mathcal{U}$ selected at random from the testing dataset, and (ii) around the profiled ribs, i.e., over a uniform partition of $n_\text{loc}=300$ locations within the output domain $Y$.}
  \label{fig:B-DeepONet-Nusselt}
\end{figure}
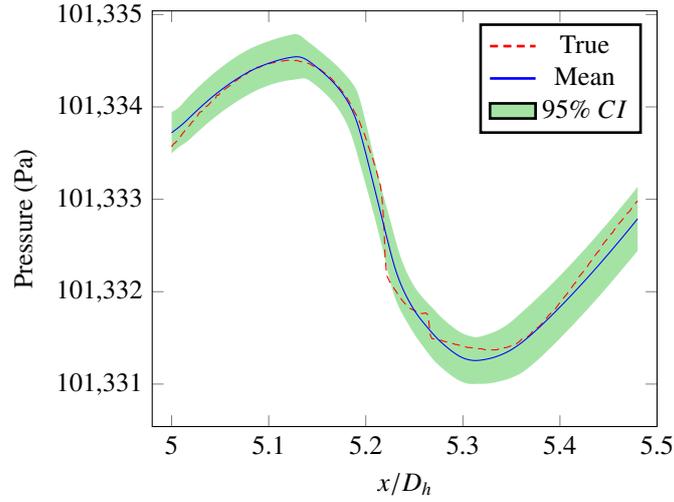
\begin{figure}[!t]
  \centering
  \begin{tikzpicture}
        \begin{axis}[width=\linewidth*0.5, xlabel={$x/D_h$}, xmin = 4.98, xmax=5.5, ymin=101330.5, ymax=101335, ylabel={Pressure (Pa)}, legend pos=north east, line width=1pt, yticklabel style={scaled ticks=false, /pgf/number format/fixed, /pgf/number format/precision=1}]  
             \addplot[name path=Pr_true,densely dashed, color=red] table{data/Pr_true.txt};
             \addlegendentry{True}
             \addplot[name path=Pr_mean, color=blue] table{data/Pr_mean.txt};
             \addlegendentry{Mean}
             \addplot[name path=Pr_lower,  fill=none, draw=none,,forget plot] table{data/Pr_lower.txt};
             \addplot[name path=Pr_upper, fill=none, draw=none,,forget plot] table{data/Pr_upper.txt};
            \definecolor{mycolor}{RGB}{75,200,75}
            \addplot[fill=mycolor, fill opacity=0.5] fill between[of=Pr_upper and Pr_lower, ];
            
            \addlegendentry{$95 \% ~ CI$}

        \end{axis}
  \end{tikzpicture}
  \caption{B-DeepONet mean prediction and confidence interval (CI) for pressure true distribution, obtained using the validated numerical approach (i) for a rib profile $u_j \in \mathcal{U}$ selected at random from the testing dataset, and (ii) around the profiled ribs, i.e., over a uniform partition of $n_\text{loc}=300$ locations within the output domain $Y$.}
  \label{fig:B-DeepONet-pressure}
\end{figure}

\section{Rib Optimization with DeepONet- and B-DeepONet-based Surrogates} \label{sec:optimization}
This section describes how to use the proposed DeepONet- or B-DeepONet-based surrogates to optimize rib profiles for internal cooling channels. Formally, the goal is to find an optimized rib profile $u^* \in \mathcal{U}$ by solving the following optimization problem:
$$
u^* = \text{arg}~\max_{u \in \mathcal{U}} J(u; G_{\theta^*}, \mathcal{D})
$$
where $J$ is the reward function. Here, we have explicitly stated the dependence of the reward function on the trained surrogate model $G_{\theta^*}$ (DeepONet or B-DeepONet) and the dataset $\mathcal{D}$.

To solve the optimization problem mentioned above, we build upon the cross-entropy method proposed by \cite{botev2013cross} and develop a gradient-free and constrained DeepONet-based (or B-DeepONet-based) cross-entropy method to seek for an optimized rib profile~$u^* \in \mathcal{U}$. Our method combines a sampling step, a constrained satisfaction step, and a selection step to efficiently explore the design space of rib profiles and reduce the computational burden of evaluating all possible profiles. We would like to emphasize that although our proposed optimization method is gradient-free, our DeepONet- and B-DeepONet-based surrogate models can also be used to aid optimization with gradient-based methods or Langevin Monte-Carlo. We plan to investigate these optimization methods in future work.

We provide details of the proposed method in Algorithm~\ref{alg:DeepONet-CEM}. The algorithm takes the following inputs: a parametric distribution $p(u;\lambda)$ of rib profiles (e.g., a Gaussian distribution) with initial parameters $\lambda_o$, the trained DeepONet or the B-DeepONet distribution obtained using the ensemble of parameters, the number of elites $n_e$, the reward function $J$ (detailed in the next section), and the number of iterations. Algorithm~\ref{alg:DeepONet-CEM} first employs the \textit{sampling} step to obtain a set of $n \gg n_e$ candidate rib profiles. Then, in the \textit{constraint satisfaction} step, the algorithm verifies whether all sampled rib profiles belong to the constraint set~$\mathcal{U}$. Finally, for the \textit{selection} step, the algorithm first evaluates the reward for all feasible rib profiles using the forward pass of DeepONet or B-DeepONet, and then selects the $n_e$ elite rib profiles with the highest reward. Using the $n_e$ elite rib profiles, Algorithm~\ref{alg:DeepONet-CEM} updates the parameters~$\lambda$ of the distribution~$p$. These steps are repeated for $T$ iterations until the optimized rib profile $u^* \sim p(u;\lambda^*)$ is returned.
\begin{algorithm}[t]
\DontPrintSemicolon
\SetAlgoLined
\textbf{Require:} a parametric distribution of rib designs~$p(u;\lambda)$, the initial parameters~$\lambda_0$, the trained DeepONet $G_{\theta^*}(\cdot)(Y_{\text{test}})$ or fitted B-DeepONet distribution $\mathcal{N}(\bar{\mu}(\cdot)(Y_{\text{test}}), \bar{\sigma}_e(\cdot)(Y_{\text{test}}))$, the number of \textit{elites}~$n_e$, the reward function~$J$, and the number of iterations~$T$.\;
Initialize $\lambda = \lambda_0$\;
\For{$t = 1,\ldots,T$}{
  sample a set of $n \gg n_e$  candidate rib designs $u_1,\ldots,u_n$ from $p(u;\lambda)$\;
  set reward $J(u_i) = - \infty$ for all infeasible designs, i.e., all $u_i \not \in \mathcal{U}$ for $i=1,\ldots,n$\;
  evaluate the reward functions for all $n' \le n$ feasible designs $J(u_1), \ldots, J(u_{n'})$\;
\textit{ \color{black} $J(u_i)$ is computed using the DeepONet forward pass $G_{\theta^*}(u_i)(Y_{\text{test}})$} \textbf{or}\; 
\textit{ \color{black} $J(u_i)$ is computed using a sample from the B-DeepONet fitted distribution $\mathcal{N}(\bar{\mu}(u_i)(Y_{\text{test}}), \bar{\sigma}_e(u_i)(Y_{\text{test}}))$}\;
select the elites $J(u_{i_1}), \ldots, J(u_{i_{n_e}})$ with the highest reward value, where $n_e<n$\;
refit the parameters $\lambda$ of the distribution $p(u;\lambda) $ to the elites $u_{i_1}, \ldots, u_{i_{n_e}} $.\;
}
  \textbf{Return:} optimized design $u^* \sim p(u; \lambda^*)$.\;
 \caption{DeepONet-based Cross-Entropy Method~\cite{botev2013cross} for Rib Optimization}
 \label{alg:DeepONet-CEM}
\end{algorithm}
\subsection{The Reward Function} \label{subsec:reward-function}
In this subsection, we describe the reward function $J$ used within the proposed optimization framework detailed in Algorithm~\ref{alg:DeepONet-CEM}. 
To enhance heat transfer and pressure drop, we propose using a continuous variation of thermal performance (also known as the thermal enhancement factor) described in~\eqref{eq:TP} as the reward function, \ie
$$
J \left(u;\theta^{Nu*},\theta^{P*} \right) = \frac{\overline{Nu}/Nu_o}{(f/f_o)^{1/3}}.
$$
In the above, we have explicitly stated the dependence of the reward function on the trained parameters of DeepONet or B-DeepONet for the Nusselt number and pressure distribution.

To compute the reward $J$, we proceed as follows: for a given rib profile candidate $u \in \mathcal{U}$, we need to compute the average Nusselt number over the channel distribution:
$$
\overline{Nu} = \frac{1}{\text{length}(Y)} \int_Y Nu(u)(y) dy.
$$
We approximate the integral above by using the forward pass of the trained DeepONet for the distribution of the Nusselt number as follows:
$$
\overline{Nu} \approx \frac{1}{n_p} \sum_{y_i \in \mathcal{P}_{\text{opt}}} G_{\theta^{Nu*}}^{Nu}(u)(y_i).
$$
Here, $\mathcal{P}_\text{opt} \subset Y$ denotes an arbitrary partition of $x$-wall locations with size $n_p$. It is worth noting that DeepONet allows for the use of random or uniformly spaced partitions. For simplicity, this paper uses a uniform partition of $n_p=300$ locations.

To compute the friction $f$, we need to first compute the pressure drop $\Delta P$. We approximate this pressure drop using the trained DeepONet forward pass as follows:
\kh{$$
\Delta P \approx G_{\theta^{P*}}^{P}(u)(y_{\min}) - G_{\theta^{P*}}^{P}(u)(y_{\max}),  
$$}
where $y_{\max}$ and $y_{\min}$ are respectively the inlet and outlet locations.

Finally, for B-DeepONet instead of using the DeepONet forward pass, we sample from the corresponding fitted Gaussian distribution $\mathcal{N}(\bar{\mu},\sigma_{e})$ of the Nusselt number or the pressure distribution
\subsection{Rib Optimization with DeepONet and B-DeepONet Numerical Results} \label{subsec:rib-optimization-numerical-results}
This subsection presents the results of rib profile optimization obtained using two different surrogate models: (i) deterministic DeepONet-based surrogates and (ii) B-DeepONet-based surrogates.

\begin{figure}[!t] 
\begin{subfigure}[!bl]{0.3\textwidth}
	\centering
	\begin{tikzpicture}[line cap=round,line join=round, scale=1]
		\draw node[anchor=north west, inner sep=0](replica){};
		\draw [line width=1pt] (0.0,0.0)-- (0.5,0.0);
		\draw [line width=1pt] (0.5,0)-- (0.5,3.2);
		\draw [line width=1pt] (0.5,3.2)-- (3.7,3.2);
		\draw [line width=0.75pt,dotted] (4.0,3.2)-- (4.5,3.2);
		\draw [line width=1pt] (3.7,3.2)-- (3.7,0.0);
		\draw [line width=1pt] (3.7,0.0)-- (4.2,0.0);
		\draw [<->,line width=0.75pt] (4.45,3.2)-- (4.45,0);
		\draw (4.5,1.75) node[anchor=north west] {\textbf{e}};
		\draw [->,line width=0.75pt] (2.35,0)-- (2.35,0.75);
		\draw [->,line width=0.75pt] (2.35,0)-- (3,0);
		\draw (2.7,0.45) node[anchor=north west] {\textbf{x}};
		\draw (2.35,1.0) node[anchor=north west] {\textbf{y}};
		\draw [fill=black] (2.35, 0) circle (2pt);
    \end{tikzpicture}
	\caption{Baseline rib profile}
	\label{subfig:Baseline}
	\end{subfigure}
\hfill
\begin{subfigure}[!br]{0.3\textwidth}
	\centering
	\begin{tikzpicture}[line cap=round,line join=round, scale=1]
        \draw node[anchor=north west, inner sep=0](replica){};
        \draw [black, xshift=0cm] plot [smooth, tension=0.7] coordinates {(-1.6, 0) (-1.60237329, 0.76314876) (-1.51766366, 1.45664907) (-1.81282335, 2.54664838) (-1.51784998, 3.11354404) (-0.60756276, 3.36902621) (0.75608037, 3.0874792) (1.43042305, 3.12836271) (1.64848274, 2.16152771) (1.59563568, 1.67688691) (1.69457642, 0.77040089) (1.6, 0)};
        \draw [<->,line width=0.75pt] (2.35,3.2)-- (2.35,0);
        \draw (2.45,1.75) node[anchor=north west] {\textbf{e}};
        \draw [->,line width=0.75pt] (0,0)-- (0,1.0);
        \draw [->,line width=0.75pt] (0,0)-- (1,0);
		\draw (0.9, 0.40) node[anchor=north west] {\textbf{x}};
		\draw (0.05, 1.1) node[anchor=north west] {\textbf{y}};
		\draw [fill=black] (0, 0) circle (2pt);
		\begin{scriptsize}
        \draw [color=red] (-1.6, 0)-- ++(-3pt,-3pt) -- ++(6pt,6pt) ++(-6pt,0) -- ++(6pt,-6pt);
        \draw [color=red] (0,3.25)-- ++(-3pt,-3pt) -- ++(6pt,6pt) ++(-6pt,0) -- ++(6pt,-6pt);
        \draw [color=red] (1.6, 0)-- ++(-3pt,-3pt) -- ++(6pt,6pt) ++(-6pt,0) -- ++(6pt,-6pt);
        \draw [fill=black] (-1.60237329, 0.76314876) ++(-3.5pt,0 pt) -- ++(3.5pt,3.5pt)--++(3.5pt,-3.5pt)--++(-3.5pt,-3.5pt)--++(-3.5pt,3.5pt);
        \draw[color=black] (-1.85,0.8) node {$1$};
        \draw [fill=black] (-1.51766366, 1.45664907) ++(-3.5pt,0 pt) -- ++(3.5pt,3.5pt)--++(3.5pt,-3.5pt)--++(-3.5pt,-3.5pt)--++(-3.5pt,3.5pt);
        \draw[color=black] (-1.85, 1.50) node {$2$};
        \draw [fill=black] (-1.81282335, 2.54664838) ++(-3.5pt,0 pt) -- ++(3.5pt,3.5pt)--++(3.5pt,-3.5pt)--++(-3.5pt,-3.5pt)--++(-3.5pt,3.5pt);
        \draw[color=black] (-2.1, 2.6) node {$3$};
        \draw [color=blue, fill=blue,shift={(-1.51784998, 3.11354404)},rotate=180] (0,0) ++(0 pt,3.75pt) -- ++(3.25pt,-5.625pt)--++(-6.5pt,0 pt) -- ++(3.25pt,5.6pt);
        \draw[color=black] (-1.75, 3.15) node {$4$};
        \draw [color=blue, fill=blue,shift={(-0.60756276, 3.36902621)},rotate=180] (0,0) ++(0 pt,3.75pt) -- ++(3.25pt,-5.625pt)--++(-6.5pt,0 pt) -- ++(3.25pt,5.6pt);
        \draw[color=black] (-0.45, 3.6) node {$5$};
        \draw [color=blue,fill=blue,shift={(0.75608037, 3.0874792)},rotate=180] (0,0) ++(0 pt,3.75pt) -- ++(3.25pt,-5.625pt)--++(-6.5pt,0 pt) -- ++(3.25pt,5.625pt);
        \draw[color=black] (0.65, 3.3) node {$6$};	
        \draw [color=blue,fill=blue,shift={(1.43042305, 3.12836271)},rotate=180] (0,0) ++(0 pt,3.75pt) -- ++(3.25pt,-5.625pt)--++(-6.5pt,0 pt) -- ++(3.25pt,5.625pt);
        \draw[color=black] (1.6, 3.3) node {$7$};		
        \draw [fill=black] (1.64848274, 2.16152771) ++(-3.5pt,0 pt) -- ++(3.5pt,3.5pt)--++(3.5pt,-3.5pt)--++(-3.5pt,-3.5pt)--++(-3.5pt,3.5pt);
        \draw[color=black] (1.9, 2.2) node {$8$};
        \draw [fill=black] (1.59563568, 1.67688691) ++(-3.5pt,0 pt) -- ++(3.5pt,3.5pt)--++(3.5pt,-3.5pt)--++(-3.5pt,-3.5pt)--++(-3.5pt,3.5pt);
        \draw[color=black] (1.85, 1.5) node {$9$};	
        \draw [fill=black] (1.69457642, 0.77040089) ++(-3.5pt,0 pt) -- ++(3.5pt,3.5pt)--++(3.5pt,-3.5pt)--++(-3.5pt,-3.5pt)--++(-3.5pt,3.5pt);
        \draw[color=black] (1.95, 0.8) node {$10$};		
      \end{scriptsize}
	\end{tikzpicture}
	\caption{Best rib profile geometry $u^* \in \mathcal{U}$ obtained by optimizing the thermal performance~$J$ using Algorithm~\ref{alg:DeepONet-CEM} with deterministic DeepONets and for $T=200$ iterations.}
	\label{fig:best-geometry}
\end{subfigure}
\hfill
\begin{subfigure}[!bl]{0.3\textwidth}
	\centering
	\begin{tikzpicture}[line cap=round,line join=round, scale=1]
        \draw node[anchor=north west, inner sep=0](replica){};
        \draw [black, xshift=0cm] plot [smooth, tension=0.5] coordinates {
        (-1.6, 0) (-1.611095280, 0.93768456) (-1.62253804, 1.58458997) (-1.79655854, 2.55398356) (-1.76688112, 2.95809473) (-0.69749272, 3.46923926) (0.7948853, 2.9893563) (1.52363743, 3.05565096) (1.6096756, 2.32726342) (1.62474926, 1.4793346) (1.64448564, 0.84074444) (1.6, 0)};
        \draw [<->,line width=0.75pt] (2.35,3.2)-- (2.35,0);
        \draw (2.45,1.75) node[anchor=north west] {\textbf{e}};
        \draw [->,line width=0.75pt] (0,0)-- (0,1.0);
        \draw [->,line width=0.75pt] (0,0)-- (1,0);
		\draw (0.9, 0.40) node[anchor=north west] {\textbf{x}};
		\draw (0.05, 1.1) node[anchor=north west] {\textbf{y}};
		\draw [fill=black] (0, 0) circle (2pt);
		\begin{scriptsize}
        \draw [color=red] (-1.6, 0)-- ++(-3pt,-3pt) -- ++(6pt,6pt) ++(-6pt,0) -- ++(6pt,-6pt);
        \draw [color=red] (0,3.25)-- ++(-3pt,-3pt) -- ++(6pt,6pt) ++(-6pt,0) -- ++(6pt,-6pt);
        \draw [color=red] (1.6, 0)-- ++(-3pt,-3pt) -- ++(6pt,6pt) ++(-6pt,0) -- ++(6pt,-6pt);
        \draw [fill=black] (-1.611095280, 0.93768456) ++(-3.5pt,0 pt) -- ++(3.5pt,3.5pt)--++(3.5pt,-3.5pt)--++(-3.5pt,-3.5pt)--++(-3.5pt,3.5pt);
        \draw[color=black] (-1.85, 0.95) node {$1$};
        \draw [fill=black] (-1.62253804, 1.58458997) ++(-3.5pt,0 pt) -- ++(3.5pt,3.5pt)--++(3.5pt,-3.5pt)--++(-3.5pt,-3.5pt)--++(-3.5pt,3.5pt);
        \draw[color=black] (-1.85, 1.50) node {$2$};
        \draw [fill=black] (-1.79655854, 2.55398356) ++(-3.5pt,0 pt) -- ++(3.5pt,3.5pt)--++(3.5pt,-3.5pt)--++(-3.5pt,-3.5pt)--++(-3.5pt,3.5pt);
        \draw[color=black] (-2.1, 2.6) node {$3$};
        \draw [color=blue, fill=blue,shift={(-1.76688112, 2.95809473)},rotate=180] (0,0) ++(0 pt,3.75pt) -- ++(3.25pt,-5.625pt)--++(-6.5pt,0 pt) -- ++(3.25pt,5.6pt);
        \draw[color=black] (-1.75, 3.2) node {$4$};
        \draw [color=blue, fill=blue,shift={(-0.69749272, 3.46923926)},rotate=180] (0,0) ++(0 pt,3.75pt) -- ++(3.25pt,-5.625pt)--++(-6.5pt,0 pt) -- ++(3.25pt,5.6pt);
        \draw[color=black] (-0.45, 3.6) node {$5$};
        \draw [color=blue,fill=blue,shift={(0.7948853, 2.9893563)},rotate=180] (0,0) ++(0 pt,3.75pt) -- ++(3.25pt,-5.625pt)--++(-6.5pt,0 pt) -- ++(3.25pt,5.625pt);
        \draw[color=black] (0.65, 3.3) node {$6$};	
        \draw [color=blue,fill=blue,shift={(1.52363743, 3.05565096)},rotate=180] (0,0) ++(0 pt,3.75pt) -- ++(3.25pt,-5.625pt)--++(-6.5pt,0 pt) -- ++(3.25pt,5.625pt);
        \draw[color=black] (1.6, 3.3) node {$7$};		
        \draw [fill=black] (1.6096756, 2.32726342) ++(-3.5pt,0 pt) -- ++(3.5pt,3.5pt)--++(3.5pt,-3.5pt)--++(-3.5pt,-3.5pt)--++(-3.5pt,3.5pt);
        \draw[color=black] (1.9, 2.2) node {$8$};
        \draw [fill=black] (1.62474926, 1.4793346) ++(-3.5pt,0 pt) -- ++(3.5pt,3.5pt)--++(3.5pt,-3.5pt)--++(-3.5pt,-3.5pt)--++(-3.5pt,3.5pt);
        \draw[color=black] (1.85, 1.5) node {$9$};	
        \draw [fill=black] (1.64448564, 0.84074444) ++(-3.5pt,0 pt) -- ++(3.5pt,3.5pt)--++(3.5pt,-3.5pt)--++(-3.5pt,-3.5pt)--++(-3.5pt,3.5pt);
        \draw[color=black] (1.95, 0.8) node {$10$};		
      \end{scriptsize}
    \end{tikzpicture}
    \caption{\kh{Best rib profile geometry $u^* \in \mathcal{U}$ obtained by optimizing the thermal performance~$J$ using Algorithm~\ref{alg:DeepONet-CEM} with Bayesian DeepONets and for $T=100$ iterations.}}
    \label{fig:Bayesian-geometry}
    \end{subfigure}
\caption{Comparison of the optimized rib profile.}
\label{NuandPrPredic}
\end{figure}
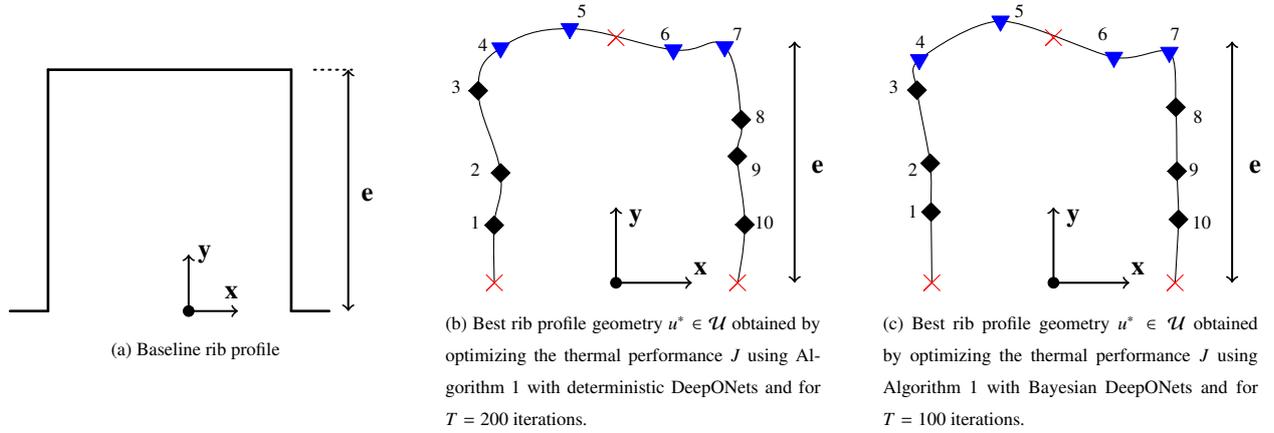

\textit{Deterministic DeepONet-based Surrogates.} We employed the Cross Entropy Method (CEM) detailed in Algorithm~\ref{alg:DeepONet-CEM} to optimize ribs using deterministic DeepONet-based surrogates. To initialize CEM, we used a normal distribution of rib designs whose initial parameters were fitted using the random rib profiles $u_j$ that we used during training. During each optimization iteration, we sampled $n=200$ rib designs and selected $n_e=20$ elites for updating the parameters of the normal distribution. We optimized the thermal performance reward $J$ using a uniform optimization partition $\mathcal{P}_{\text{opt}}$ of $n_p=300$ $x$-wall points. Figure~\ref{fig:reward} illustrates how Algorithm~\ref{alg:DeepONet-CEM} improves thermal performance $J$ until it reaches a local maximum. Additionally, Figure~\ref{fig:best-geometry} presents the optimized design~$u^* \in \mathcal{U}$ obtained from the proposed optimization Algorithm.
\begin{figure}[!t]
  \centering
  \begin{tikzpicture}
        \begin{axis}[width=\linewidth*0.5, xlabel={Iteration~$t$},  ylabel={Reward~$J$}, legend pos=south east, line width=1pt, yticklabel style={scaled ticks=false, /pgf/number format/fixed, /pgf/number format/precision=1}]  
             \addplot[name path=reward, solid, color=blue] table{data/reward_curve.txt};
        \end{axis}
  \end{tikzpicture}
  \caption{Reward (thermal performance) $J$ trajectory obtained by employing Algorithm~\ref{alg:DeepONet-CEM} using trained deterministic DeepONets for pressure and Nusselt number distributions over $T = 200$ iterations.}
  \label{fig:reward}
\end{figure}
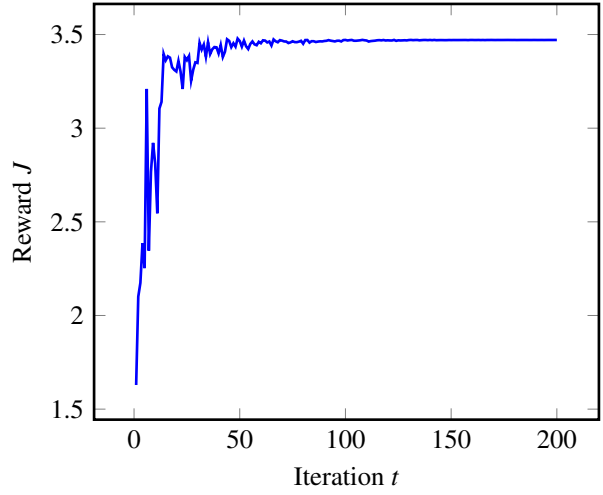

\kh{\textit{Bayesian DeepONet-based Surrogates.} We also used the CEM method (Algorithm~\ref{alg:DeepONet-CEM}) to optimize rib profiles using Bayesian DeepONet-based surrogates. Figure~\ref{fig:Bayesian-geometry} shows the best rib profile geometry $u^* \in \mathcal{U}$ obtained by optimizing the thermal performance~$J$ using Algorithm~\ref{alg:DeepONet-CEM} with Bayesian DeepONets, for $T=100$ iterations.}

\kh{Figure~\ref{fig:rib-profile-comparison} compares the Nusselt number ratio and pressure drop for three different rib profiles: (i) square rib profile, (ii) optimized rib profile using Algorithm~\ref{alg:DeepONet-CEM} and the deterministic DeepONet-based surrogate, and (iii) Bayesian rib profile using Algorithm~\ref{alg:DeepONet-CEM} and the Bayesian DeepONet-based surrogate. The results illustrate improved pressure drop and Nusselt number ratio for the rib profiles optimized using the proposed DeepONet- and B-DeepONet-based surrogates compared to square rib profile. In particular, the percentage change of enhanced Nusselt number ratio and pressure drop for the proposed surrogates is depicted in Figure~\ref{fig:percentage-change}.}

\usetikzlibrary {spy}
\begin{figure}[!b] 
	\begin{subfigure}[!br]{0.45\textwidth}
	\begin{tikzpicture}[spy using outlines={circle, magnification=3,size=1.6cm, connect spies}]
		\begin{axis}[width=\linewidth,
			xmin = 0.0,xmax=10,	xlabel={x/D$_h$},	ymin=0, ymax=6, ylabel={Nu$_{ave-local}$/Nu$_o$},	legend pos=north east,xtick={0,1,...,10}, ytick={0,1,2,...,6},legend cell align={left},]
			\addplot [smooth, black, mark=square*, mark size=1.2pt] 
			table{data/Nuwithoutrib.txt}; 	\label{Nusseltex}; 
            \addlegendentry[font=\small, align=left]{Square rib profile}
			\addplot [smooth, blue, mark=triangle*, mark size=1.2pt] 
            table{data/NuwithoutRibOptimized.txt}; 	\label{Nusseltlocal};
            \addlegendentry[font=\small, align=left]{Optimized rib profile}
            \addplot [smooth, cyan, mark=star, mark size=1.2pt] 
            table{data/BayesianNu.txt}; 	\label{BayesianNu}; 
            \addlegendentry[font=\small, align=left]{Bayesian rib profile}
            \addplot+ [only marks, mark=square*, mark size=1.5pt,mark options={draw=red,fill=red}]
			table{data/Rib.txt};	\label{RibLocation1}; 
            \addlegendentry[font=\small, align=left]{Rib Locations}	
            \coordinate (spypoint) at (axis cs:5.48, 2.6);
            \coordinate (magnifyglass) at (axis cs:3, 1.3);
		\end{axis}
        \spy on (spypoint) in node at (magnifyglass);
	\end{tikzpicture}
	\caption{Local Nusselt number ratio comparison between optimized and square rib profiled channel}
	\label{sub-fig:NuratioOp1}
\end{subfigure}
\hfill 
	\begin{subfigure}[!br]{0.45\textwidth}
	\begin{tikzpicture}[spy using outlines={circle, magnification=3,size=2cm, connect spies}]
		\begin{axis}[width=\linewidth, xmin = 0.0,xmax=10,	xlabel={x/D$_h$}, ymin=101320, ymax=101360,  ylabel={Pressure (Pa)},	legend pos=north east,xtick={0,1,...,10}, line width=1pt, yticklabel style={scaled ticks=false, /pgf/number format/fixed, /pgf/number format/precision=1}, legend cell align={left},]
            \addplot [line width=0.75pt, black, mark=square*, mark size=1.5pt, mark indices={1,323}] 
            table{data/PressureSquare.txt}; 	\label{Pressure1}; 
            \addlegendentry[font=\small,]{Square rib profile}
            \addplot [line width=0.75pt, blue, mark=triangle*, mark size=1.5pt, mark indices={1,323}] 
            table{data/PressureOptimized.txt}; 	\label{Pressure2}; 
            \addlegendentry[font=\small,]{Optimized rib profile}
            \addplot [line width=0.75pt, cyan, mark=star, mark size=1.5pt, mark indices={1,323}] 
            table{data/BayesianPressure.txt}; 	\label{Pressure3}; 
            \addlegendentry[font=\small,]{Bayesian rib profile}
            \addplot+ [only marks, mark=square*, mark size=1.5pt,mark options={draw=red,fill=red}]
		  table{data/RibPressure.txt};	\label{RibLocation1}; 
            \addlegendentry[font=\small,]{Rib Locations}	
            \coordinate (spypoint) at (axis cs:5.13, 101332);
            \coordinate (magnifyglass) at (axis cs:1.85, 101332);
		\end{axis}
        \spy on (spypoint) in node at (magnifyglass);
	\end{tikzpicture}
	\caption{Pressure drop comparison between the optimized and the square rib profiled channel}
	\label{sub-fig:NuratioOp2}
\end{subfigure}
\caption{\kh{Comparison of the Nusselt number ratio and pressure drop for three different rib profiles: (i) square rib profile, (ii) optimized rib profile using Algorithm \ref{alg:DeepONet-CEM} and the deterministic DeepONet-based surrogate, and (iii) Bayesian rib profile using Algorithm \ref{alg:DeepONet-CEM} and the Bayesian DeepONet-based surrogate.}}
\label{fig:rib-profile-comparison}
\end{figure}
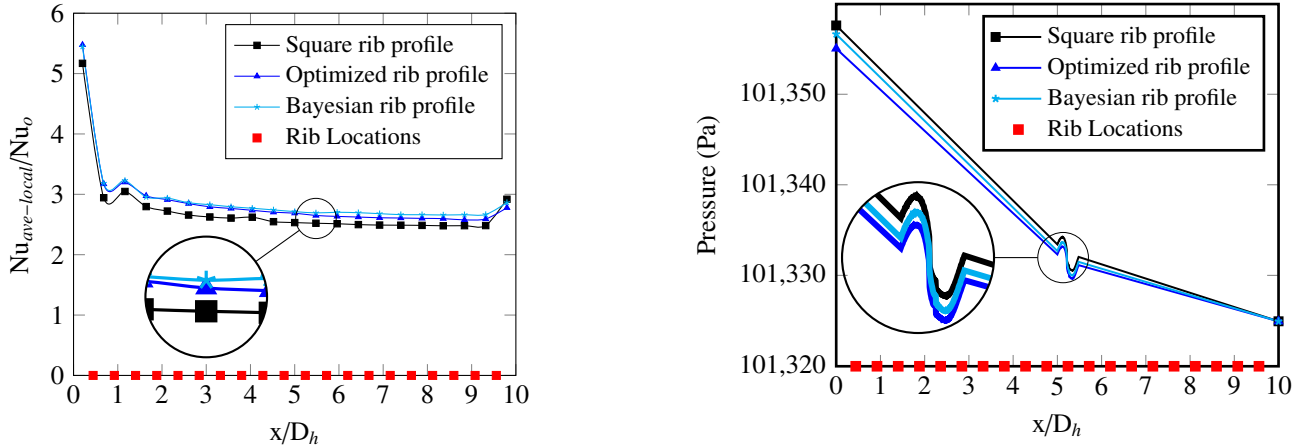

\begin{figure}[!b]
\centering
\begin{tikzpicture}
\begin{axis} 
    [ybar = 2pt, bar width = 10pt, symbolic x coords={Nu, Pressure}, domain=1:2,
    ylabel={Percentage \%}, xlabel={Based on channel averaged Nusselt number and pressure drop}, 
    ymin=-8, ymax = 8, ytick={-8,-6,...,8},
    xtick=data,
    x=2.5cm,
    enlarge x limits=0.4]
\addplot [draw = blue, line width = .4mm, fill = blue] coordinates {(Nu,5.19) (Pressure,-7.79)};  
\addplot [draw = cyan, line width = .4mm, fill = cyan] coordinates {(Nu, 7.12) (Pressure, -2.93)}; 
\legend{DeepONet, B-DeepONet}
\end{axis}
\end{tikzpicture}
\caption{\kh{Percentage change in Nusselt Number and pressure drop when comparing optimized rib profiles using DeepONet- and B-DeepONet-based surrogates to the reference square profile.}}
\label{fig:percentage-change}
\end{figure}

\section{Discussion} \label{sec:discussion}
This section discusses our surrogate models and optimization results, and provides a preamble to our future work.

\subsection{Results} 
We provide now a summay of our results. (i) This paper developed an effective DeepONet framework for predicting the distribution of pressure and Nusselt number for arbitrary random rib profiles. Specifically, we used a relatively small dataset of less than $N=275$ random rib profiles to train DeepONets, which achieved mean $L_1$ trajectory errors of less than 6\% for pressure distribution and 8\% for Nusselt number (see Table~\ref{table:DeepONet-results}) for rib profiles not included in the training dataset. It is worth noting that these error results can be easily improved by increasing the size of the training dataset. However, obtaining arbitrarily large training datasets may be prohibitively expensive in more realistic scenarios.

(ii) To handle even smaller training datasets, we designed a Bayesian DeepONet (B-DeepONet) that can quantify epistemic uncertainty. The proposed B-DeepONet can create credible intervals (see Figures~\ref{fig:B-DeepONet-pressure} and \ref{fig:B-DeepONet-Nusselt}) that capture true pressure and Nusselt number distributions for rib profiles not included in the training dataset. As a result, B-DeepONet enables reliable pressure and Nusselt number predictions within credible intervals with high probability. This helps protect against adversarial rib designs during optimization.

(iii) Finally, we used DeepONet and B-DeepONet to optimize rib profiles and improve thermal performance. To achieve this goal, we implemented a gradient-free and constrained optimization strategy detailed in Algorithm~\ref{alg:DeepONet-CEM}. As shown in Figures~\ref{fig:reward}, the proposed optimization framework that uses DeepONet- and B-DeepONet-based surrogates improved thermal performance (reward) and produced optimized rib profiles (see Figures~\ref{fig:best-geometry} \kh{and \ref{fig:Bayesian-geometry}}) that we validated using numerical methods \kh{(see Figures~\ref{fig:rib-profile-comparison} and \ref{fig:percentage-change})}. 

\subsection{Future Work}
\textit{Rib design with uncertainty quantification.} It is worth noting that the proposed DeepONets are differentiable models. As such, in our future work, we plan to develop gradient-based optimization methods to improve thermal performance through rib design. Additionally, we intend to implement the Langevin and replica-exchange MCMC sampling methods that we developed in~\cite{moya2023deeponet,lin2023b}. These methods aim to identify not only the best optimized rib geometry, but also the distributions of designs. Such a distribution will allow us to quantify the uncertainty, enabling us to better explore the rib design landscape and guide the collection of new high-fidelity data for training DeepONets.

\textit{Surrogate models and multi-fidelity DeepONets.} Collecting high-fidelity data to train DeepONet models, which take random rib profiles as input and output pressure and Nusselt number distributions, can be prohibitively expensive. Therefore, designing a multi-fidelity DeepONet-based surrogate model is part of our future work. This model will be trained using vast amounts of low-fidelity data (e.g., a dataset with lower resolution trajectories) and a few selected high-fidelity trajectories.

\textit{Towards a continual Bayesian multi-fidelity framework for rib design.} Our ultimate goal is to create a continual learning Bayesian DeepONet surrogate model for rib optimization. Our plan is to first train a DeepONet model with a large amount of low-fidelity data, which will serve as a prior model. Next, we will design and train a B-DeepONet model using the prior model and high-fidelity data. B-DeepONet will enable us to quantify uncertainty and identify underexplored rib design regions. Thus, by continuously interacting with a high-fidelity simulator, B-DeepONet and uncertainty quantification will allow us to improve the surrogate for rib design using high-fidelity data from the identified underexplored regions.
\section{Conclusion} \label{sec:conclusion}
In this paper, we designed surrogate models with uncertainty quantification capabilities to effectively improve the thermal performance of rib-turbulated internal cooling channels. To construct the surrogates, we designed (i) a Deep Operator Network (DeepONet) framework for deterministic prediction and (ii) a Bayesian DeepONet (B-DeepONet) framework for uncertainty quantification. Our proposed DeepONets take arbitrary continuous rib geometries with control points as inputs, outputting detailed information and credible intervals about the distribution of pressure and heat transfer around the profiled ribs. To train the DeepONets, we collected datasets by simulating a 2D rib-roughened internal cooling channel and adjusting the input rib geometry by randomly adjusting the control points within predefined constraints. Finally, we demonstrated the performance of our DeepONet-based surrogate models with uncertainty quantification by incorporating them into a constrained, gradient-free optimization problem, which enhances the thermal performance of the rib-turbulated internal cooling channel.

\section*{Acknowledgments}
IS gratefully acknowledge the support of Lillian Gilbreth Postdoctoral Fellowships from Purdue University's College of Engineering.

GL gratefully acknowledges the support of the National Science Foundation (DMS-1555072, DMS-2053746, and DMS-2134209), and U.S. Department of Energy (DOE) Office of Science Advanced Scientific Computing Research program DE-SC0021142 and DE-SC0023161.

\bibliographystyle{elsarticle-num} 
\bibliography{refpaper.bib, refbook.bib}
\end{document}